%% file: Belloni_Motta.tex
\documentclass[graybox,usenatbib]{svmult}

\usepackage{mathptmx}       
\usepackage{helvet}         
\usepackage{courier}        
\usepackage{type1cm}        
\usepackage{rotating}
   
\usepackage{makeidx}         
\usepackage{graphicx}        
\usepackage{multicol}        
\usepackage[bottom]{footmisc}

\include{def_bibtex}

\makeindex             

\begin{document}

\title*{Transient Black Hole Binaries}
\author{Tomaso M. Belloni and Sara E. Motta}
\institute{Tomaso M. Belloni \at INAF - Osservatorio Astronomico di Brera, via E. Bianchi 46, I-23807 Merate, Italy \email{tomaso.belloni@brera.inaf.it}
\and Sara E. Motta \at University of Oxford, Department of Physics, Astrophysics, Denys Wilkinson Building, Keble Road, Oxford OX1 3RH, UK \email{sara.motta@physics.ox.ac.uk}}
%
%
\maketitle

\abstract{
The last two decades have seen a great improvement in our understanding of the complex phenomenology observed in transient black-hole binary systems, especially thanks to the activity of the Rossi X-Ray Timing Explorer satellite, complemented by observations from many other X-ray observatories and ground-based radio, optical and infrared facilities.\\
Accretion alone cannot describe accurately the intricate behavior associated with black-hole transients and it is now clear that the role played by different kinds of (often massive) outflows seen at different phases of the outburst evolution of these systems is as fundamental as the one played by the accretion process itself.  
The spectral-timing states originally identified in the X-rays and fundamentally based on the observed effect of accretion, have acquired new importance as they now allow to describe within a coherent picture the phenomenology observed at other wavelength, where the effects of ejection processes are most evident. \\
With a particular focus on the phenomenology seen in the X-ray band, we review the current state-of-the-art of our knowledge of black hole transients, describing the accretion-ejection connection at play during outbursts through the evolution of the observed spectral-timing properties.
Although we mainly concentrate on the observational aspects of the global X-ray transient picture, we also provide physical insight to it by reviewing (when available) the theoretical explanations and models proposed to explain the observed phenomenology.
}

\include{Introduction}

\include{X-ray_emission}

		\include{Energy_spectra}

		\include{Fast_time_variability}

		\include{Long-term}

\include{Radio_IR_emission}

		\include{relativistic_jets}

		
		\include{Accretion-ejection}

			\include{Correlations}
			
\include{wind_outflows}

\include{the_full_picture}

\section{Conclusions}\label{sec:5}
\include{conclusions}



\bibliographystyle{plain}
\bibliography{biblio}

\tableofcontents

\label{lastpage}
\end{document}

%% file: Introduction.tex
\section{Introduction}\label{sec:1}

Black Hole (BH) Binaries (hereafter BHBs) are binary systems consisting of a non-collapsed star and a black hole. They provide a unique laboratory for the study not only of accretion of matter onto a compact object, but also of the strongly curved spacetime in the vicinity of a black hole, where the effects of General Relativity in a strong gravitational field cannot be ignored and can be studied (see \cite{Psaltis2008}).

The first BHB, Cygnus X-1, was discovered in the early years of X-ray astronomy with instruments on board sounding rockets. It became the first candidate for systems hosting a black hole when optical observations revealed its binary nature and allowed the estimate of the mass of the compact object (see Chapter 3 in this book). Its X-ray properties included strong variability on short time scales and a very hard energy spectrum. Cyg X-1 is a persistent system, which we know now are rather rare. Most BHBs are transient, difficult to discover in absence of a wide-field X-ray instrument. The first such system, A~0620-00 was discovered in 1975 with the Ariel V satellite \cite{Elvis1975} and in 1986 it became a much stronger candidate for hosting a black hole, with a minimum mass of 3.2 M$_\odot$ for the compact object \cite{McClintock1986a}. A few additional persistent systems were identified, notably the two in the LMC and the galactic source GX 339-4, but it was only with the Japanese satellite Ginga that new transients were discovered and followed, thanks to the availability of an all-sky monitor and of a large-area detector. Systems like GX 339-4 and Cyg X-1 were known to undergo transitions in their flux, spectral and timing properties, which had led to the definition of a high (flux) and low state (see e.g. \cite{Tananbaum1972}). The detailed study of several observations, in particular of the transient GX 1124-683 (also known as Nova Muscae 1991),  GX 339-4 and Cyg X-1 led to the identification of complex spectral/timing properties that in turn led to the definition of additional states (see \cite{Miyamoto1992,Miyamoto1993,Belloni1996,Belloni1997}). In particular, the first spectral studies with accretion disc models opened the way to techniques to measure the inner radius of the disc. In the timing domain Quasi-Periodic Oscillations (QPO) were discovered. These features yield discrete frequencies which constitute a very direct measurement of time scales in the system, which are associated to the accretion phenomenon and/or to effects of General Relativity. A full review of the observational status after Ginga can be found in \cite{Tanaka1995}), where essentially all known information on BHBs at the time is recorded.

The situation changed dramatically starting from the second half of the nineties. At the end of 1995, the Rossi X-ray Timing Explorer (RXTE) was launched and opened a new window onto phenomena of fast variability from neutron-star and black-hole binaries. Its All-Sky Monitor (ASM) allowed the timely discovery of outbursts of old and new transients and the large-area Proportional Counter Array (PCA), operated in a very flexible way, accumulated high signal data for timing analysis and spectral analysis at moderate spectral resolution. A few years later, the availability of high spectral resolution instruments on board Chandra and XMM-Newton opened the way to perform detailed spectral studies, in particular on relativistically-skewed emission lines. Overall, there was a real explosion of information and now, after the demise of RXTE and with the availability of further missions such as INTEGRAL, Swift, Suzaku and NuSTAR, it is possible to combine archival data and new observations to obtain new insight on the emission properties of BHBs. This allows us to understand much better the process of accretion onto a black hole and new exciting results are possible that were not even imaginable only a few years ago.
The long term evolution of the X-ray emission, in particular for transients, is now clear in its phenomenology and provides a solid framework upon which to study detailed spectral and variability properties. 
Direct effects of General Relativity are studied and measured through the analysis of broad-band spectra, emission lines and QPOs (see below and Chapter 3), the properties of continuum emission are interpreted with physical models and global models for different states are being proposed.
 
 At the same time, the discovery of relativistic jets from BHBs \cite{Mirabel1992,Mirabel1994} (see also \cite{Gallo2010,Fender2014}) opened a new path to the study of accretion. The difference in time scale and count statistics with Active Galactic Nuclei, where jets were known since much earlier, allows the study of the connection between accretion onto the compact object and ejection of relativistic jets, introducing a new window to understand the physics of accretion and to unveil the mechanism responsible for the ejections (see \cite{Gallo2010,Fender2014}). Moreover, in the past few years the existence of powerful non-relativistic winds has been discovered in a number of objects (see \cite{Fender2014}). These winds appear to be flowing in a direction parallel to the accretion disc, as opposed to jets, and to be mutually exclusive with the jets. The full picture of an accreting black hole is now much different from that of only a couple of decades ago, when observations in the X-ray band alone and only at low energy resolution made us miss very important components (jets and winds) which can be energetically very important. Jets and outflows are the subject of Chapter 3.
 
In addition to the comparison with Active Galactic Nuclei, the supermassive counterparts of stellar-mass accreting black holes, the strong similarities between BHBs and other systems at sub-galactic scale are being studied and exploited. Neutron-star binaries share many properties in the X-ray  band and, when their magnetic field is sufficiently low, also in the radio, where relativistic jets have been observed \cite{Munoz-Darias2014,Migliari2006}. Binaries in external galaxies are also accessible with current instrumentation, although naturally at lower count rates, and in particular the enigmatic Ultra-Luminous X-ray Sources (ULXs) are being studied in order to understand whether they contain intermediate-mass black holes or are super-Eddington counterparts of our galactic objects (see \cite{Feng2011}). In this respect, a detailed comparison with BHBs is instrumental to unveil this mystery.

In this review, we aim at presenting the current state of the art of BHB research, concentrating on the properties of accretion and ejection, while the Chapter 3 addresses the issue of measurement of physical quantities of the central object. The sheer amount of information available today makes it impossible to cover all aspects. Many specific and adidtional details can also be found e.g. in \cite{Belloni2011,Fender2012,Fender2014,Kylafis2015}.

%% file: X-ray_emission.tex
\section{X-ray emission}\label{sec:2}

In BHBs, the X-ray emission originates from the inner regions of the accretion flow and, possibly, from the base of the relativistic jets \cite{Markoff2010}. Even restricting ourselves to the two original source states, low/hard and high/soft, the details of the emission appear to be very complex (see e.g. \cite{Gilfanov2010}). Both the energy spectra and the fast variability contain multiple components, which vary in a correlated fashion. While the details of these properties are given in the next sections, it is important to understand the regularities that exist in the time evolution of the observables. Before the mid 90's, only a few sources were available and the coverage was not sufficient to study these aspects in sufficient detail; only with RXTE it has been possible to find a general scheme for the characterisation of source states. Even before then, it was clear that a simple classification of the observed properties on the basis of energy spectra alone was not possible and fast variability had to be included in the picture. In this sense, the notion of `spectral states' does not carry any meaning.

While a few BHBs are persistent sources, like the archetypal system Cyg X-1 (always accreting at a high rate and emitting luminosities above $10^{37}$ erg/s) most of them are of transient nature. They spend most of the time in a low accretion regime (L$_X < 10^{33}$ erg/s), where observations are still limited by the low number of counts (see \cite{Plotkin2015} and references therein). With a recurrence period that varies between several months and decades, the accretion rate onto the central objects increases by orders of magnitude and the sources go into outburst for a time that can range from a few days to, more commonly, several months (one peculiar object, GRS 1915+105 is at present active since 23 years). Their X-ray luminosity increases, peaks and then decreases and can roughly be adopted as a proxy for accretion rate, while the detailed properties of the energy spectra and fast variability change, at times in a very abrupt way. It was only at the beginning of the last decade decade however that a coherent picture emerged, which can be applied to most systems \cite{Homan2001,Homan2005a}.

\begin{figure}[t]
\includegraphics[scale=0.46]{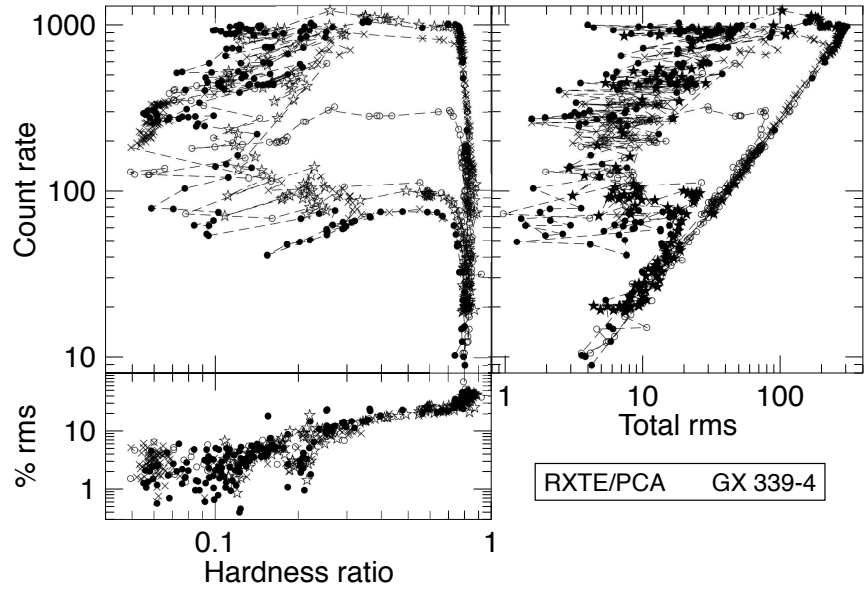}
\caption{The three main diagrams for the representation of the X-ray evolution of black-hole binaries for three outbursts of GX 339-4 as observed with RossiXTE. Top left: Hardness-Intensity Diagram (HID), top right: Rms-Intensity Diagram (RID), bottom left: Hardness-Rms Diagram: (HRD).
}
\label{fig:diagrams}
\end{figure}

Outbursts of different systems and even multiple outbursts from the same object have time evolutions which can differ considerably (see Belloni 2010). However, when the evolution of an outburst is represented in a Hardness-Intensity Diagram (HID), strong regularities emerge. A HID is equivalent to a hardness-magnitude diagram in the optical band: on the abscissa is the ratio of counts in two separate bands (hard/soft), which gives a rough indication of the hardness of the energy spectrum, and on the ordinate the total count rate over a broad energy band, a proxy for luminosity and accretion rate \cite{Homan2001,Homan2005,Belloni2010,Belloni2011}. As such, the diagram is source dependent (because of interstellar absorption) and instrument-dependent, but it is extremely useful to follow BHB outbursts. An extension to an independent form of the diagram (with flux ratio between components and total flux respectively) has been proposed \cite{Dunn2010}: this has the advantage of containing physical quantities, but the disadvantage of being insensitive to small changes when one component dominates, besides being of course model-dependent. An example of a HID based on RXTE/PCA data for the best known system, GX 339-4, is shown in the top left panel of Fig. \ref{fig:diagrams}. Here, four outbursts are plotted (2002, 2004, 2007, 2010). The general evolution is the same: a `q' shaped diagram traveled counterclockwise from the bottom right (faint and hard). Quiescence cannot be included as the RXTE/PCA instrument was not sensitive to faint observations, but we know that the hard branch softens as flux decreases (see \cite{ArmasPadilla2011} and references therein). From the HID, one can easily identify the two historical states as the two ``vertical'' branches. The hard branch extending to the stem of the `q' corresponds to the Low/Hard State (LHS), which is observed at the start and at the end of an outburst only, never in the middle. The branch is not really vertical as the logarithmic axis suggests, but there is a marked softening as the source brightens (see e.g. \cite{Motta2009a}). The left branch corresponds to the High/Soft State (HSS). The scatter of the points is magnified by the log scale and contains some excursions back to other states (see below) as well as intrinsic variability of the hard spectral component. The transition between these two states takes place at two different flux levels. At high flux, the source moves from LHS to HSS and at low flux it returns to the LHS, completing a hysteresis cycle (originally identified by \cite{Miyamoto1995}, see also \cite{Maccarone2003,Kylafis2015}).
Two additional diagrams have been proven useful for following the evolution of an outburst and identify source states. The first is the HRD (Hardness-Rms Diagram, bottom left panel in Fig. \ref{fig:diagrams}), where the Y axis contains the fractional rms variability integrated over a broad range of frequencies (see \cite{Belloni2010}). Remarkably, no hysteresis is observed in this diagram: at each hardness corresponds a single value of fractional rms, regardless of the flux. The second is the RID (Rms-Intensity Diagram, right panel in Fig. \ref{fig:diagrams}), the third combination of the same three observables. Here the hysteresis, as expected, is clearly present and the location of specific states is rather precise (see \cite{Munoz-Darias2011}).
The central part of the diagram identifies two additional states, the Hard Intermediate State (HIMS) and Soft Intermediate State (SIMS). The transition between states corresponds to precise values in hardness, obviously source and instrument dependent. The identification of these thresholds is based on the properties of fast variability and/or changes in the multi wavelength relations. For a precise determination of the different states we refer the reader to \cite{Belloni2011} and \cite{Munoz-Darias2011}, here we include a brief summary:

\begin{itemize}

\item Low-Hard State (LHS). The LHS has only been observed in the first and last stages of an outburst. In some cases, the first RXTE/PCA observations found a source already in a softer state, but the initial LHS might have been too fast to observe. It corresponds to the right branch in the HID and to the straight diagonal line in the RID (see Fig. \ref{fig:diagrams}). It is characterised by large variability (around 40\% fractional rms in the case of GX 339-4) and a hard spectrum (see below). In transients, it is only observed at the start and end of outbursts (although at times the start LHS is missed altogether) and is the most common state for the persistent system Cyg X-1). The variability is in the form of broad-band noise made of a few components whose characteristic frequencies increase with luminosity (and decreasing hardness).

\item Hard Intermediate State (HIMS). In the HID, this state corresponds to a large part of the area between the LHS and the HSS, covering the horizontal tracks, both at high and low flux. This state appears after the initial LHS and reappears before the source goes back to the LHS at the end of the outburst. In addition, secondary transitions to and from it can be observed (see below). The softening compared to the LHS is due to two effects: the appearance in the observational range of flux from the thermal disc and the steepening of the hard component (see below and \cite{Motta2009a}). The fast variability is an extension of that of the LHS, with characteristic frequencies increasing and total fractional rms decreasing (see the HRD and the RID in Fig .\ref{fig:diagrams}). Type-C QPOs are present (see below).

\item Soft Intermediate State (SIMS). The energy spectrum is slightly softer than the HIMS, putting this state to the left of the HIMS in the HID. In the HID and RID these points are not immediately identifiable, but can be seen in the HRD (bottom panel in Fig. \ref{fig:diagrams}) as a cloud of points at a lower rms than the main branch (around hardness 0.2). While the energy spectrum below 10 keV is very similar to that of the softer HIMS points, at high energies the spectrum unlike the HIMS does not show a significant high-energy cutoff  \cite{Motta2009a}. The identifying feature of this state is the disappearance of the band-limited noise components in the power density spectrum, replaced by a weaker power law component (hence the lower fractional rms) and the appearance of a marked type-B QPO.

\item High-Soft State (HSS). The spectrum is soft, dominated by an optically thick accretion disc, variability is low. Occasional low-frequency QPOs can be detected, identified with type-C (see \cite{Motta2012}). This state is reached from the intermediate states and left through the intermediate states. 

Although for different sources diagrams can look different in the HID in Fig. \ref{fig:diagrams} the sequence of states from quiescence to quiescence is the following: LHS -- HIMS -- SIMS -- HSS -- (minor transitions to and from HIMS and SIMS) -- HSS -- SIMS -- HIMS -- LHS.  Some transient remained in the LHS throughout the outburst, a few showed failed transitions in the sense that the LHS-HIMS sequence was not followed by a transition to the softer states (SIMS and HSS). The few bright persistent sources show a reduced number of states. Cyg X-1 is found mostly in the LHS with transition to the HIMS and possibly all the way to the HSS. LMC X-1 is always in the HSS. LMC X-3 is mostly in the HSS, with brief transitions to the LHS (no strong evidence of intermediate states).

The three diagrams presented above and the state classification are a firm basis upon which to base detailed spectral and timing analysis. One important transition is the HIMS--SIMS one, which appears to be associated to the ejection of relativistic ballistic jets (the crossing of the ``jet line,'' see below).

\end{itemize}

%
%
%

%% file: Energy_spectra.tex
\subsection{Energy spectra}\label{subsec:2.1}

\subsubsection{The truncated disc model}

Long term X-ray light curves, X-ray spectra, the rapid
X-ray variability and the radio jet behaviour have been shown to be consistent with the so-called \textit{truncated disc model}. According to this model, at low luminosities a Shakura-Sunyaev optically thick, geometrically thin accretion disc is truncated at a certain (variable) radius and coexists with a hot, optically thin, geometrically thick accretion flow, which replaces the region between the inner edge of the disc and the innermost stable orbit.
Neutron stars are also consistent with the same description (\cite{Munoz-Darias2014}), but with an additional component due to their surface, giving implicit evidence for the event horizon in black holes.


At low luminosities, the optically thick, geometrically thin disc is truncated at very large radii, being
replaced (probably through evaporation, see e.g. \cite{Meyer1994}) in the inner regions by a hot, inner flow which might also act as the launching site of the jet.
Only a few photons from the disc  illuminate the flow at this stage, therefore, Compton cooling
of the electrons is rather inefficient compared to the heating coming from collisions with
protons. The ratio of power in the electrons to that in the seed photons illuminating
them - L$_{h}$/L$_{s}$ - , is the major parameter (together with the optical depth
of the plasma) which determines the shape of a thermal Comptonization
spectrum (e.g. \cite{Haardt1993}). Physically, L$_{h}$/L$_{s}$ sets the energy
balance between heating and cooling, and hence the electron temperature. 
In the hard state, the relative lack of seed photons illuminating the hot
inner flow produces hard thermal Comptonized
spectra (with L$_{h}$/L$_{s}$ $>>$ 1),  roughly characterized by a power law in the 5--20 keV
band with photon index 1.5 $< \Gamma  <$ 2 (where the photon spectrum N(E) $\propto$
E$^{−Γ}$). Hard spectra of this kind are typical of the LHS.

As the disc moves progressively inwards, it increasingly extends underneath the hot inner
flow so that there are more seed photons intercepted by the flow, decreasing
 L$_{h}$/L$_{s}$.  Therefore, the decrease in disc truncation radius leads to softer spectra, as well as higher
frequencies  in the power spectra and a faster jet. This results in spectra that are a combination of the hard spectral component described above and the soft spectral component typical of the soft state (see below), i.e. a standard geometrically thin, optically thick accretion disc with a progressively smaller inner radius. These spectra are observed during the (usually) short-lived HIMS and SIMS.

When the truncation radius reaches the innermost stable orbit, the hot flow is thought to collapse into a Shakura-Sunyaev disc and  dramatic changes in both the spectral and time domain are seen. This includes a significant decrease in radio flux, as well as the major hard-to-soft spectral transition seen in BHs.  The dramatic increase in disc flux due to
the presence of the inner disc marks the hard-soft state transition (\cite{Esin1997}), and also means
that any remaining electrons which gain energy outside of the optically thick
disc material are subject to much stronger Compton cooling (L$_{h}$/L$_{s}$ is now $\leq$1) . This
results in much softer Comptonized spectra. Thus the soft
state is characterized by a strong disc and soft tail, roughly described by
a power law index of photon index $\Gamma \geq$ 2, extending out beyond
500 keV (\cite{Gierlinski1999}). This tail, differently from the hard tail observed in the LHS,  is not produced by thermal Comptonization. In order to extend to 500 keV and beyond, the spectrum should be produced in a region with rather small optical depth and high temperature. However, these conditions would result in a bumpy spectrum, with individual Compton scattering
orders separated, in contrast with the observed smooth powerlaw-like tail. Such a spectrum can be instead produced by  Compton scattering on a non-thermal electron population, where the index is set predominantly by the shape of the electron distribution rather
than L$_{h}$/L$_{s}$.
Soft spectra such as the ones described here are typically see during the HSS.



As noted in \cite{Done2007}, even though there are no observations which unambiguously conflict with the
truncated disc models, there exist a tremendous amount of data which
can be fit within this geometry (which includes, besides energy spectra, rapid variability characteristics and jet properties). 
While the truncated
disc model is indeed currently a very simplified version of what must be
a more complex reality, nonetheless the range of data it can qualitatively explain gives confidence that it captures the
essence of the main spectral states.

\subsubsection{Alternative geometries}

Alternative geometries that include an untruncated disc and (mostly) isotropic
source emission have significant problems in matching the observed features
of the hard-state spectra. 
\begin{itemize}

\item The shape of the spectral conitnuum observed in the hard state rules out slab corona
models as the spectra all peak at high energies. These spectra are possible only if the  luminosity in seed photons within
the X-ray region is less than that in the hot electrons (i.e. L$_{h}$/L$_{s}$ $>>$ 5), while in a slab geometry only quite small L$_{h}$/L$_{s}$  can be obtained (a disc
extending underneath an isotropically radiating hot electron region would
intercept around half the Comptonized emission in a slab corona geometry, see \cite{Haardt1993}). 

\item A patchy corona allows part of the reprocessed flux to escape without re–illuminating the hot electron
region, so it can produce the required hard spectra. A patchy corona
also allows the reflected flux to escape along with the reprocessed flux, 
resulting in a strong reflection spectrum for very hard spectra, in direct conflict with the
observations (see e.g. \cite{Malzac2005}). 


\item Models where the X-rays are produced directly in the jet were proposed
by \cite{Markoff2001}. These produce the hard X-rays by synchrotron
emission from the high-energy extension of the same non-thermal electron
distribution which gives rise to the radio emission. However, the observed
shape of the high-energy cutoff in the hard state is very sharp, and cannot
be easily reproduced by synchrotron models \cite{Zdziarski2003}.
However, there are now composite models where the X-rays are from
Comptonization by thermal electrons at the base of the jet (which resides in a hot flow at the centre of a truncated disc), while the radio is from non-thermal electrons accelerated up the jet (\cite{Markwardt2005}). This model practically  converges onto the truncated disc model, though with some additional weak beaming of the hard X-rays. 

\item One last alternative to the truncated disc model is represented by the magnetized accretion-ejection model of \cite{Ferreira2006}, which has a disc-inner jet structure similar to that by \cite{Markwardt2005}, though here the inner, optically thick
disc is still present down to the last stable orbit, but with properties very different from the standard Shakura-Sunyaev disc. The transition radius between this jet dominated disc and the standard accretion disc is variable, producing the range of behaviour seen in the hard state spectra in a similar way to the truncated disc model.

\end{itemize}

From what has been said above, all currently viable models for the hard state converge on a geometry
where the standard disc extends down only to some radius larger than the last stable orbit, with the properties of the flow abruptly changing at this point.

%% file: Fast_time_variability.tex
\subsection{Fast time variability}\label{subsec:2.2}

Fast time variability is an important characteristic of BHBs and a key ingredient for understanding the physical processes in these systems. Fast (aperiodic and quasi-periodic) variability is generally studied through the inspection of power density spectra (PDS; \cite{VdK1989}). Most of the power spectral components in the PDS of BHBs are broad and can take the form of a wide power distribution over several decades of frequency or of a more localized peak (Quasi-Periodic Oscillations, QPOs). 

QPOs were discovered several decades ago in the  X-ray flux emitted from accreting neutron stars and have since been observed in many BHB systems (see e.g. \cite{Motta2015}).  It is now clear that QPOs are a common characteristic of accreting BHs and they have been observed also in neutron stars (NS) binaries (e.g. \cite{VdK1989}, \cite{Homan2002a}, \cite{Belloni2007}), in cataclysmic variables (see e.g. \cite{Patterson1977}), in the so-called \textit{ultra luminous X-ray sources} (possibly hosting intermediate-mass BHs or super-Eddington accreting NS, e.g.  \cite{Strohmayer2003a}, \cite{Bachetti2014}) and even in Active Galactic Nuclei (AGNs, e.g. \cite{Gierlinski2008}, \cite{Middleton2010}). 

\subsubsection{Low Frequency QPOs}

Low-frequency QPOs (LFQPOs) with frequencies ranging from a few mHz to $\sim$30 Hz are a common feature in almost all transient BHBs and were already found in several sources with \emph{Ginga} and divided into different classes  (see e.g. \cite{Miyamoto1991} for the case of GX 339-4 and \cite{Takizawa1997} for the case of GS 1124-68). Observations performed with the Rossi X-ray Timing Explorer (RXTE) have led to an extraordinary progress in our knowledge of the properties of variability in BHBs (see \cite{VDK2006}, \cite{Remillard2006}, \cite{Belloni2011}): it was only after RXTE was launched that LFQPOs were detected in most observed BHBs (see \cite{VDK2004}. 

Three main types of LFQPOs, dubbed types A, B, and C, originally identified in the PDS of XTE J1550-564 (see \cite{Wijnands1999}; \cite{Homan2001}; \cite{Remillard2002}), have been seen in several sources.  
The different types of QPOs are currently identified on the basis of their intrinsic properties (mainly centroid frequency and width, but energy dependence and phase lags as well), of the underlying broad-band noise components (noise shape and total variability level) and of the relations among these quantities. 

\begin{figure}[t]
\includegraphics[width=1\textwidth]{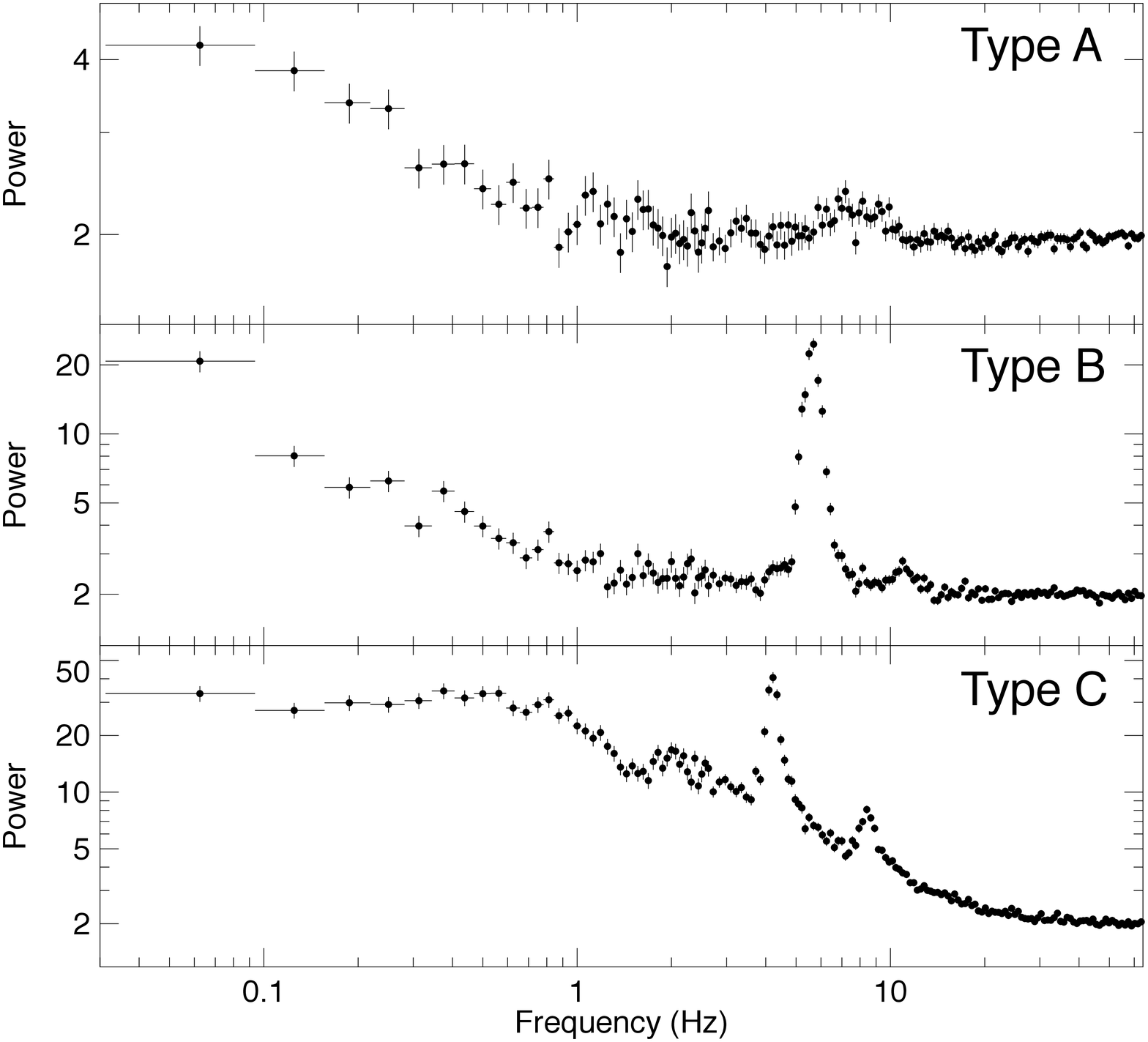}
\caption{Examples of type A, B and C QPOs from our GX 339-4 observations. The contribution of the Poisson noise was not subtracted. Adapted from \cite{Motta2011a}.}
\label{fig:LFQPOs}
\end{figure}

\begin{itemize}

\item Type-A QPOs (Fig.  \ref{fig:LFQPOs}, top panel) are characterized by a weak (few percent rms) and broad ($\nu/\Delta\nu$ $\leq$3) peak around 6-8 Hz. Neither a subharmonic nor a second harmonic are usually present (possibly because of the width of the fundamental peak), whereas a very low-amplitude red noise is associated with these QPOs. Originally, these LFQPOs were dubbed \textit{type A-II} by \cite{Homan2001}. LFQPOs dubbed \textit{type A-I} (\cite{Wijnands1999}) were strong, broad and associated with a very low-amplitude red noise. A shoulder on the right-hand side of this QPO was clearly visible and interpreted as a very broadened second
harmonic peak. \cite{Casella2005} showed that this \textit{type A-I} LFQPOs should be classified as a Type-B QPOs. Type-A QPOs usually appear in the HSS, just after the transition from the HIMS.

\item Type-B QPOs (Fig. \ref{fig:LFQPOs}, middle panel) are characterized by a relatively strong
(4\% rms) and narrow ($\nu/\Delta\nu$ $\geq$6) peak, which is found in a narrow range of centroid frequencies around 6 Hz or 1-3 Hz (\cite{Motta2011}). A weak red noise (few percent rms or less) is detected at very low frequencies ($\leq$0.1 Hz). A weak second harmonic is often present, sometimes together with a subharmonic peak. In a few cases, the subharmonic peak is higher and narrower, resulting in a \textit{cathedral-like} QPO shape (see \cite{Casella2004}). 
Rapid transitions in which type B LFQPOs appear/disappear are often observed in some sources (e.g. \cite{Nespoli2003}). These transitions are difficult to resolve, as they take place on a timescale shorter than a few seconds. The presence of type-B QPOs essentially defines the SIMS. 

\item Type-C QPOs (Fig.  \ref{fig:LFQPOs}, bottom panel) are characterized by a
strong (up to 20\% rms), narrow ($\nu/\Delta\nu$ $\geq$10) and variable peak (its centroid frequency and intensity varying by several percent in a few days; see, e.g., \cite{Motta2015}) at frequencies 0.1-15 Hz, superimposed onto a flat-top noise  that steepens above a frequency comparable to that of the QPO. A subharmonic and a second harmonic peak are often seen, and sometimes even a third harmonic peak.  The total (QPO plus flat-top noise) fractional rms variability can be as high as 40\%. The frequency of the type-C QPOs correlates both with the flat-top noise break-frequency (\cite{Wijnands1999a} and with the characteristic frequency of some broad components seen in the PDS at higher frequency ($>$20Hz, see \cite{Psaltis1999}). Type-C QPOs are usually observed during the bright end of the LHS and during the HIMS. In some sources (see eg \cite{Motta2012}, \cite{Motta2015}), type-C QPOs can be seen also in the HSS, where they show frequencies that can reach $\sim$30 Hz.
\end{itemize}

\subsubsection{Models for LFQPOs}

Despite LFQPOs being known for several decades, their origin is still not understood and there is no consensus about their physical nature. However, the study of LFQPOs provides an indirect way to explore the accretion flow around black holes (and neutron stars).

The existing models that attempt to explain the origin of LFQPOs are generally based on two different mechanisms: instabilities and  geometrical effects. In the latter case, the physical process typically invoked is precession.

\cite{Titarchuk2004} proposed the so called \textit{transition layer model}, where type-C QPOs are the result of viscous magneto-acoustic oscillations of a spherical bounded transition layer, formed by matter from the accretion disc adjusting to the sub-keplerian boundary conditions near the central compact object.\\
\cite{Cabanac2010} proposed a model to explain simultaneously type-C QPOs and the associated broad band noise. Magneto-acoustic waves propagating within the corona makes it oscillate, causing a modulation in the efficiency of the Comptonization process on the embedded photons. This should produce both the type-C QPOs (thanks to a resonance effect) and the noise that comes with them.\\
\cite{Tagger1999} proposed a model based on the \textit{accretion ejection instability} (AEI), according to which a spiral density wave in the disc, driven by magnetic stresses, becomes unstable by exchanging angular momentum with a Rossby vortex. This instability forms low azimuthal wavenumbers, standing spiral patterns which would be the origin of LFQPOs. \cite{Varni`ere2002} and \cite{Varni`ere2012} suggested that all the tree types of QPOs (A, B and C) can be produced through the AEI in three different regimes: non-relativistic (type-C), relativistic (type-A, where the AEI coexist with the Rossby Wave Instability (RWI), see \cite{Tagger1999}) and during the transition between the two regimes (type-B QPOs). 

\cite{Stella1998} proposed the so called \textit{relativistic precession model} (RPM) to explain the origin and the behaviour of a type of LFQPO (the so called horizontal-branch oscillation) and of two high-frequency QPOs (the so-called kHz QPOs) in NS X-ray binaries, as the result of the nodal precession, periastron precession and keplerian motion, respectively, of a self-luminous blob of material in the accretion flow around the compact object. This model was later extended to BHs (\cite{Stella1999} and \cite{Motta2014a} recently showed that the RPM provides a good explanation for both type-C QPOs and high-frequency QPO (see below) in at least two BH systems. 

\cite{Ingram2009} proposed a model based on the relativistic precession as predicted by the theory of General Relativity that attempts to explain type-C QPOs and their associated noise. 
This model requires a cool optically thick, geometrically thin accretion disc  (\cite{Shakura1973}) truncated at some radius, filled by a hot, geometrically thick accretion flow. This geometry is known as \textit{truncated disc model} (\cite{Esin1997}, \cite{Poutanen1997}). 
In this framework, type-C QPOs arise from the Lense-Thirring precession of a radially extended section of the hot inner flow that modulates the X-ray flux through a combination of self-occultation, projected area and relativistic effects that become stronger with inclination (see \cite{Ingram2009}).  
The broad-band noise associated with type-C QPOs, instead, would arise from variations in mass accretion rate from the outer regions of the accretion flow that propagate towards the central compact object, modulating the variations from the inner regions and, consequently, modulating also the radiation in an inclination-independent manner (see \cite{Ingram2013}). 

\subsubsection{High Frequency QPOs}

Among the most important discoveries of RXTE is the detection of  the so-called kHz QPOs
in neutron star binaries (see \cite{VDK2006}). This result opened a window onto high-frequency
phenomena in BHBs. The first observations of the very bright system GRS
1915+105 led to the discovery of a transient oscillation at $\sim$67 Hz (\cite{Morgan1997}), the first high-frequency QPO (HFQPO) in a BHB. Since then, sixteen years of RXTE observations have yielded only
a handful of detections in other sources, although GRS 1915+105 seems to be an
exception, with a remarkably high number of detected high-frequency QPOs (see e.g. \cite{Belloni2012}).

\begin{figure}[t]
\includegraphics[width=1\textwidth]{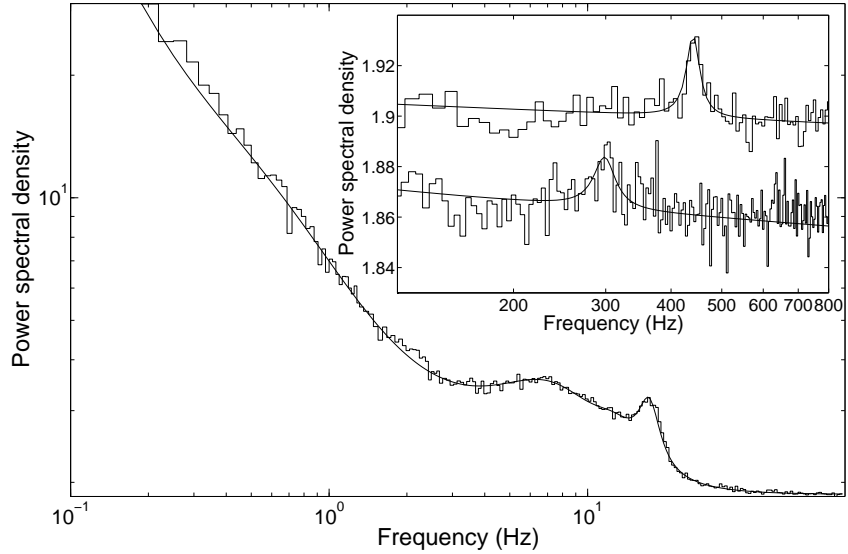}
\caption{Power spectrum of GRO J1655-40 displaying three simultaneous QPO peaks: the type C at $\sim$18 Hz, upper and lower high frequency QPO at $\sim$300 and $\sim$450 Hz, respectively. (Fig. 2 from \cite{Motta2014}).}
\label{fig:HFQPOs}
\end{figure}

The properties of the few confirmed HFQPOs (\cite{Belloni2012}) can be summarized as follow:

\begin{itemize}

\item They appear only in observations at high flux/accretion rate. This is
at least partly due to a selection effect, but not all high-flux observations lead to
the detection of a HFQPO, all else being equal, indicating that the properties of
these oscillations can vary substantially even when all other observables do not
change.

\item They can be observed as single or double peaks. Only one source, GRS J1655-40 (see Fig. \ref{fig:HFQPOs}, showed two clear simultaneous peaks
(\cite{Strohmayer2001}, \cite{Motta2014}), while all the others only showed single peaks, sometimes at different frequencies (see Tab. 1 in \cite{Belloni2014}. 
In XTE J1550-564, the two detected peaks (\cite{Remillard2002}) have been detected simultaneously after averaging a number of observations, but the lower one with a 2.3 $\sigma$ significance
 (\cite{Miller2001}). \cite{Mendez2013},
on the basis of their phase lags, suggested that the two detected peaks might be the
same physical signal at two different frequencies. H 1743-322 showed a clear HFQPO with a weak second simultaneous peak (\cite{Homan2005}).
A systematic analysis of the data from GRS 1915+105 (\cite{Belloni2013}) led to the detection of 51 HFQPOs, most of which at a centroid frequency between 63
and 71 Hz. All detections corresponded to a very limited
range in spectral parameters, as measured through hardness ratios. Additional peaks at 27, 34 and 41 Hz were discovered by \cite{Strohmayer2001a}, \cite{Belloni2001} and \cite{Belloni2013}. The most recent HFQPO discovered, in IGR
J17091-3624, is consistent with the average frequency of the 67 Hz QPO in GRS
1915+105 (\cite{Altamirano2012}).

\item Typical fractional rms for HFQPOs are 0.5-6\% increasing steeply with energy, in
the case of GRS 1915+105 reaching more than 19\% at 20-40 keV (see right panel
in Fig. 6 of \cite{Morgan1997}). Quality factors Q\footnote{Q is defined as the
ratio between centroid frequency and FWHM of the QPO peak.} are around 5 for
the lower peak and 10 for the upper. In GRS 1915+105, a typical Q of $\sim$20 is
observed, but values as low as 5 and as high as 30 are observed.

\item Time lags of HFQPOs have been studied for four sources (\cite{Mendez2013}). The
lag spectra of the 67 Hz QPO in GRS 1915+105 and IGR J170913624 and of the
450 Hz QPO in GRO J1655-40 are hard (hard photons variations lag soft photons
variations), while those of the 35 Hz QPO in GRS 1915+105 are soft. The 300 Hz
QPO in GRO J1655-40 and both HFQPOs in XTE J1550-564 are consistent with
zero (suggesting that the two HFQPOs in XTE J1550-564 are the same feature seen at different frequencies). 

\item For three sources, GRO J1655-40, XTE J1550-564 and XTE J1743-322, the two
observed frequencies are close to being in a 3:2 ratio (\cite{Strohmayer2001}, \cite{Remillard2002}, \cite{Remillard2006}), which has led to a family of models, known as  \textit{resonance models} (see e.g. \cite{Abramowicz2001}). For GRS 1915+105 the 67 Hz and 41 Hz QPOs,
observed simultaneously, are roughly in 5:3 ratio. The 27 Hz would correspond to
2 in this sequence.

\end{itemize}

\subsubsection{Models for HFQPOs}\label{sec:hfqpos_models}

Many models have been proposed to describe HFQPOs of BHBs, all involving in some form the predictions of the Theory of General relativity. 

The relativistic precession model (RPM), was originally proposed by \cite{Stella1998}, \cite{Stella1999}  to explain the origin and the behaviour of the LFQPO and kHz QPOs in NS X-ray binaries and later extended to BHs (\cite{Stella1999a},\cite{Motta2014},\cite{Motta2014a}). The RPM associates three types of QPOs observable in the PDS of accreting compact objects  to a combination of the fundamental frequencies of particle motion. The nodal precession frequency (or Lense-Thirring frequency) is associated with Type-C QPOs LFQPOs, while the periastron precession frequency and the orbital frequency are associated with the lower and upper HFQPO, respectively (or to the lower and upper kHz QPO in the case of NSs). 
The relativistic precession model has been proposed in two other versions. In \cite{Bursa2005} it is assumed that radiation is modulated by the vertical oscillations of a slightly eccentric fluid slender torus formed close to the ISCO. Stuchlik and collaborators proposed a further version of the relativistic precession model, that has been studied in many papers by this group. Here the model is related to the warped-disc oscillations discussed by \cite{Kato2004} (see below).

The the warped disc model proposed by \cite{Kato2004} and \cite{Kato2004a} states that the HFQPOs are resonantly excited by specific disc deformations – warps. The model was generalized to include precession of the warped disc in \cite{Kato2005} and spin-induced perturbations were included in  \cite{Kato2005a}.

\cite{Abramowicz2001} and \cite{Kluzniak2001} introduced the nonlinear \textit{resonance model}, which was later   studied extensively by them as well as by other authors. This model is based on the assumption that non-linear 1:2, 1:3 or 2:3 resonance between orbital and radial epicyclic motion could produce the HFQPOs observed in both BH and NS binaries. Later on, \cite{Abramowicz2004} proposed another version of this model, called  the Keplerian non-linear resonance model, where the resonance occurs between the radial epicyclic frequency and the orbital frequency instead of between the radial epicyclic frequency and the vertical frequency. 
These \textit{resonance models}  successfully explain
black hole QPOs with frequency ratio consistent with 2:3 or 1:2 (see Sect. 3.2). As a
given resonance condition is verified only at a fixed radius in the disc, the QPO frequencies
are expected to remain constant, or jump from one resonance to another.

%
%
%
%
%
%
%
%
%
%
%
%
%
%


%% file: Long-term.tex
\subsection{Long-term time evolution}\label{subsec:2.3}


	After having defined and discussed the source states in terms of their spectral and timing properties along the HID/HRD/RID diagrams, it is useful to examine the evolution of sources along the HID (and in parallel the other diagrams). A short description can be found in \cite{Fender2012}, together with an animation.
	
\begin{figure}[t]
\includegraphics[scale=0.46]{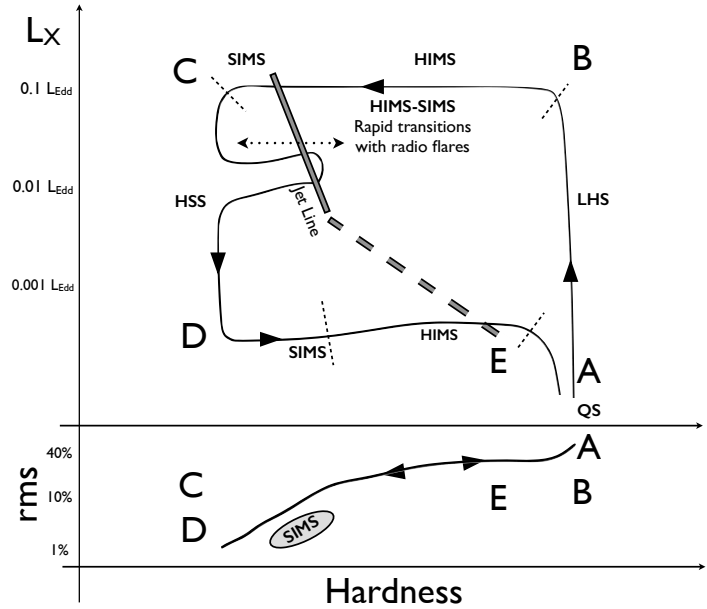}
\caption{Generic HID and HRD of a black-hole binary outburst. The letters refer to main locations described in the text. (Fig. 1 from \cite{Kylafis2015}).}
\label{fig:evolution}
\end{figure}

	BHBs spend most of the time in a ``quiescent'' state, where the accretion rate reaching the central regions of the accretion flow and the black hole is very low, typically lower than $10^{-5} L_{Edd}$ \cite{Plotkin2013,Plotkin2015}, indicated in Fig. \ref{fig:evolution} as QS but actually located below the extension of the Y axis. Here the energy spectra indicate that the spectrum deviates from that of the LHS, which hardens as flux decreases. Going down to quiescence, at a level around $10^{-2}L_{Edd}$ the measured power-law index starts increasing from 1.5-1.6 until leveling off around 2 at $10^{-5} L_{Edd}$ \cite{Wu2008,Plotkin2013}). 

	The branch from A to B in Fig. \ref{fig:evolution} corresponds to the LHS and is traveled on time scales which can be as long as months but also so short that the LHS is not observed in pointed observations made on the day following the first alert. We do not know directly that the A-B branch was traversed, but we have never seen evidence of the contrary. There are a few cases of systems which never leave the LHS and return to quiescence after having reached a peak (see e.g. \cite{Brocksopp2010} and references therein), not necessarily at low flux \cite{Oosterbroek1997}). Along this vertical branch, the characteristic frequencies of the strong noise components observed in the PDS increase, while the total fractional rms variability decreases.
	
	Then at B the HIMS is entered. The precise time of the transition can be identified with the changes of timing properties (see \cite{Belloni2005a} and \cite{Munoz-Darias2011}) or through a change in the low-energy properties as was observed in GX 339-4 (\cite{Homan2005}). The duration of the HIMS, when the source softens from B to the SIMS transition can be less than a day up to two weeks. A few sources, after entering the HIMS, did not proceed to a further transition and returned to the LHS and then to quiescence, failing to reach a full transition (\cite{Brocksopp2004,Capitanio2009,Soleri2013}). The characteristic frequencies continue to increase and the total fractional rms to decrease, while an evident type-C QPO appears, also with increasing frequency as the source softens. 
	
	If the outburst does not fail, a transition to the SIMS is made. While the energy spectrum below $\sim$10 keV is similar to that of the softest HIMS observations, with a hard-component photon index around 2.4, the SIMS is characterized by marked differences in the PDS. In particular, the disappearing of the type-C QPO and the appearing of a type-B QPO at a different frequency are relatively easy to observe. This transition has been seen to take place on a time scale of a few seconds \cite{Casella2004}.	Multiple back and forth transitions have also been observed, on time scales of days to weeks but also down to minutes \cite{Nespoli2003,Casella2004,Belloni2005a}. The evolution of the high-energy ($>10$ keV)spectral component also changes abruptly here.
	Around, but not exactly coincident to, this transition, fast discrete relativistic jets are launched, observed either as resolved moving radio spots or as bright radio flares (see \cite{Fender2004}). Although it would be tempting to causally associate the disappearance of timing components in the PDS to the ejection of the plasma responsible for them, it has been shown that in some case the ejection starts a few days {\it before} the transition \cite{Fender2009}.
	
	The position of the further transition to the HSS is not easy to identify, as the presence of weak type-A QPOs is not always easy to ascertain. The HSS, after possible back transition to the previous intermediate states, is rather stable and can last for months. The luminosity, which in many systems peaks in the HIMS/SIMS (but see below), decreases, most likely because of a more or less steady decrease in mass accretion rate. A low level of aperiodic time variability is detected, in the form of a power law component in the PDS, with a total fractional rms around 1-2\%.
	
	At a luminosity level well below that of the early HIMS, a new transition takes place (at D). From here on, time is reversed, first the SIMS and then the HIMS are observed, after which the LHS is reached, after which the outburst ends and the source goes back to its quiescent level. However, notice that at low luminosity the thermal disc contribution is softer, which means that HID points at the same hardness do not correspond to the same energy spectrum. Indeed, the photon index of the hard component at the  return SIMS is around 2.1, comparable to that at the start of the first HIMS (point B) \cite{Stiele2011}.
	
	The transition between LHS-HIMS-SIMS-HSS at high flux does not necessarily coincide to the maximum in accretion rate, as it is assumed in the previous description. Indeed, in a few cases the accretion rate continued to increase after the HSS is reached. If this happens, in the HID the source moves up from C moving to the right (see \cite{Belloni2010,Motta2012}). Both the energy spectrum and the PDS become rather complex \cite{Motta2012,Done2006}).

    There are sources for which the diagram appears more complex than this (see examples in \cite{Belloni2010,Fender2009}), as well as sources which are observed when already in a bright HSS. The latter must have come from quiescence and the data do not exclude that the LHS-HIMS-SIMS branches were followed, only on a much shorter time scale, of the order of the day.	
    
    The HID path outlined in Fig. \ref{fig:evolution} shows a clear evidence of hysteresis, as the return path is different from the forward path. In other words, the luminosity (or accretion rate) level at which the hard-to-soft transition takes place is higher than that of the reverse transition \cite{Maccarone2003}. For the prototypical source GX 339-4, which had several outbursts, there is evidence that the higher the first level, the lower the second. A correlation has been found between waiting time from the previous outburst and the hard X-ray peak, which corresponds to the hard-to-soft transition level (see \cite{Wu2010a}), but this does not specifically address the issue of hysteresis. The same effect is observed in neutron-star transients \cite{Maccarone2003,Munoz-Darias2014} and implies that luminosity (and hence accretion rate) cannot be an absolute proxy for state. In other words, LHS and HSS can be observed over the same range of luminosities, although the transition from hard to soft does take place at the highest LHS flux (see also \cite{Maccarone2003}. While most generic interpretations for the observed hysteresis invoke the presence of a second parameter in addition to accretion rate, since accretion rate determines the movement along the diagram, but the LHS-HSS transition can take place at different accretion rate levels, \cite{Kylafis2015} associate the hysteresis with the system's memory, in the same way that magnetic hysteresis works. However, what sets the transition level and whether a transition will take place or not is still an open problem.
    It is interesting to note that while the upper path can take place at very different luminosities, the range spanned by the lower path is much more reduced, around a few \% of the Eddington luminosity (see \cite{Maccarone2003,Kalemci2013}).
    Adding complication, \cite{Munoz-Darias2013} showed that there is a systematic difference in the shape of the HID between high- and low-inclination sources.
    
    A similar hysteresis diagram has been observed in a white dwarf binary, the dwarf nova SS Cyg, using optical emission as soft band and X-ray emission as hard band for the production of the HID \cite{Koerding2008}. A radio flare was also detected in correspondence to the hard-to-soft transition. Although the comparison is interesting, it has to be noted that in a white-dwarf binary the optical/UV emission originates from the accretion disc and the X-ray emission from the boundary layer between disc and surface. In general sense, what we are observing from these systems is a transition from optically thin to optically thick emission when the density has increased and a reverse transition when the density has decreased, an effect which is also valid for the boundary layer of a dwarf nova(see \cite{Kuulkers2006}). The details of the physics of the systems and the transitions do not necessarily need to be the same.
    
    As described in Sect. 2.1.1 and highlighted in more detail by \cite{Kylafis2015}, the general outburst evolution (including the basic diagrams, energy spectra and fast time variability) at zero level can be interpreted as a simple evolution of the transition radius within the truncated disc model (see also Sect. \ref{subsubsec:3.2.1}).


		

%% file: Radio_IR_emission.tex
\section{Radio/IR emission}\label{sec:3}

While the process of accretion onto stellar-mass black holes has been studied, mainly in the X-rays, since the beginning of X-ray astronomy, it is only relatively recently that observations at longer wavelengths, notably infrared and radio, have led to the discovery of additional processes such as the ejection of relativistic jets and of wind outflows. It is now clear that without a broad-band perspective, the complete and complex picture of an accreting compact object in a binary system is impossible to understand. Below, we review the main observational points that in the past decade have led to a major change in perspective.

%% file: relativistic_jets.tex
\subsection{Radio jets}\label{subsec:3.1}

\begin{figure}[t]
	\includegraphics[scale=1.26]{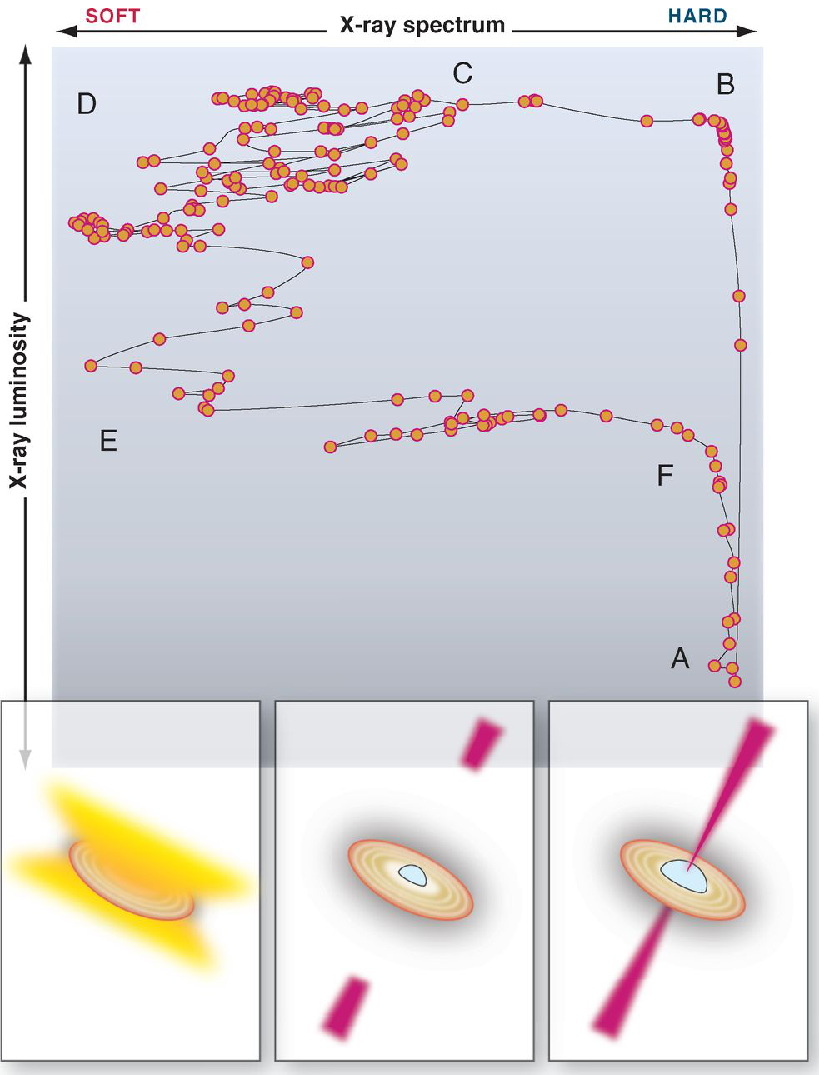}
	Figura 5: \caption{Schematic evolution of a black-hole binary along the HID (top panel). The three bottom panels show the configuration of the system along the top branch: where red represents radio jets and yellow wind outflows. From \cite{Fender2012}. Reprinted with permission from AAAS.}
	\label{fig:fb121}
\end{figure}

Although radio emission from black-hole binaries had been detected a long time before (see e.g. Tananbaum et al. 1971), it was only in the early nineties that coordinated radio campaigns were started, following the discovery of relativistic jets in the radio band. The current observational picture is rather clear and it correlates with the position of the source in the HID (see above). Figure \ref{fig:fb121} shows a sketch of a HID with an indication of the different types of outflow observed.

\begin{description}
	\item[from below A to B] BHBs spend most of their time in the quiescent state (well below point A in Fig. \ref{fig:fb121}), accessible only by the most powerful focusing X-ray telescopes. The X-ray luminosity is $< 10^{33}$erg/s. Weak flat spectrum radio emission has been detected in this state. As the source brightens in the LHS, the radio luminosity also increases, while the radio spectrum remains flat. In this state, a compact jet has been resolved in a few cases in the radio band (see e.g. )\cite{Stirling2001}). The radio spectrum is flat, consistent with self-absorbed synchrotron emission and extends up to the near infrared, while the polarization level is low. The X-ray flux and radio flux show a strong positive non-linear correlation (see Sect. \ref{subsubsec:3.2.2}). The observational data are interpreted with the presence of a compact jet emitting in the radio through synchrotron and moving outwards with moderately relativistic speed ($\Gamma < 2$, see \cite{Fender2004}).
	
	\item[from B to C] Point B marks the start of the HIMS. In addition to changes in the rapid variability (total fractional rms), a good marker for this transition is the breakdown of the correlation between radio-IR flux and X-ray flux (see below and \cite{Homan2005,Fender2004}). As the transition to the SIMS is approached, the radio emission generally decreases, with oscillations and small flares \cite{Miller-Jones2006}, while there is indication that the radio spectrum starts steepening \cite{Fender2004}. All data indicate that the inner part of the jet undergoes changes in its physical properties.
	
	\item[from C to D] As outlined above, the transition between HIMS and SIMS is marked by rapid changes in the properties of fast variability, in particular a drop in fractional rms and a switch between type-C and type-B QPOs. Around (but not precisely in correspondence to) the transition, the jet properties change drastically. One or more large-amplitude flares are observed in the radio (see \cite{Fender2009}). When major superluminal ejections are observed, the ejection time can also be traced back to a time close to the transition. Unfortunately, the lack of precise correspondence of the time of jet ejection and that of HIMS-SIMS transition indicates that there cannot be a causal connection. There is indication that these jets have a larger Lorentz factor than the steady jets in the LHS and HIMS, which led to the idea that what is observed is due to internal shocks caused by faster jets hitting the slower components (see \cite{Fender2004}).
	
    \item[from D to E] This branch marks the HSS. Until now no radio emission that can be attributed to the central source (and not from ejecta) has been detected down to upper limits of $>$300 times that of LHS sources at the same X-ray flux (see \cite{Russell2011a}). Outflows in the form of winds are observed (see \ref{sec:4}). During this state accretion rate decreases on a long time scale (weeks to months), leading to the vertical track in the HID. However, at high flux there can be multiple transitions to the SIMS/HIMS and back, as measured through the changes in timing properties \cite{Belloni2005a,Casella2004,Belloni2010,Fender2012}. Although coverage at lower energies is sparse, the observations are consistent with the presence of weaker radio flares corresponding to the transitions (see e.g. \cite{Brocksopp2002,Casella2004}.
    
    \item[from E to F] On the return branch from soft to hard, as we have seen above the reverse track is followed, through the SIMS and the HIMS, on a time scale comparable to that of the upper track. Observations of several transitions from multiple systems have shown that the compact radio jet is re-formed {\it not} in correspondence to the SIMS-HIMS transition, but when the system has reached the LHS, a delay of several days (see \cite{Kalemci2013,Kalemci2014} and references therein. At the time of radio re-appearance, secondary maxima have been seen in the optical-infrared band, which have been attributed to direct jet emission, although alternative models exist \cite{Kalemci2013}.

\end{description}

%% file: Correlations.tex
\subsubsection{Correlations}\label{subsubsec:3.2.2}

\begin{figure}[t]
	\includegraphics[scale=0.65]{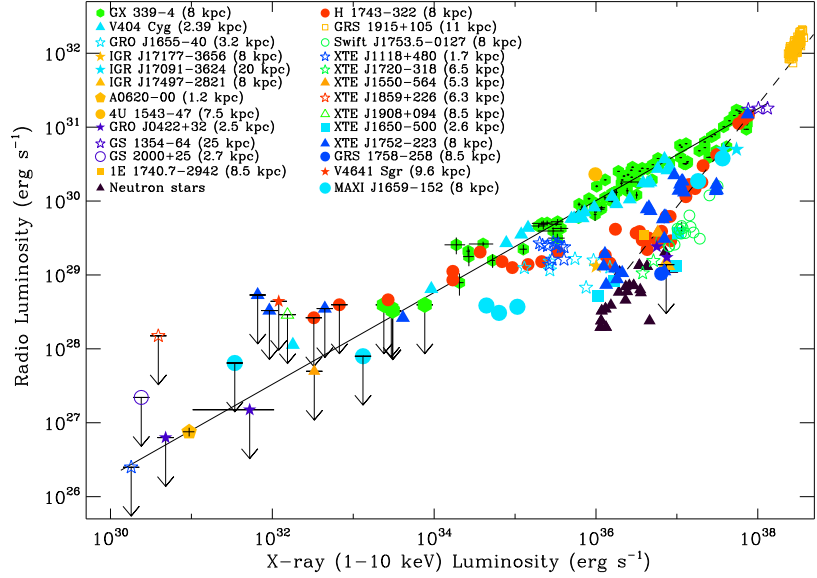}
	\caption{Correlation between radio and X-ray luminosity for BHBs in the LHS and in quiescence (Fig. 9 from \cite{Corbel2013}).}
	\label{fig:corbel}
\end{figure}

The radio and optical/IR emission from BHBs is attributed to jets (for outflows, see Sect. \ref{sec:4}) and is  closely connected to the accretion properties as measured at higher energies. In particular, the low-energy flux from the compact jet in the LHS shows a strong non-linear correlation with X-ray flux, modeled with a power law with index $\alpha\sim$0.6-0.7 (Fig. \ref{fig:corbel}, see \cite{Corbel2013,Gallo2012}, first discovered in GX 339-4 \cite{Corbel2003} then extended to other systems \cite{Gallo2003}). Ignoring the points which appear to follow a different correlation (see below) the scatter in the relation is rather small. Under the assumption that the X-ray flux originates from accretion and therefore is not subject to relativistic beaming like the radio emission from a jet, from the scatter around the correlation \cite{Fender2004} have estimated the jet Lorentz factor to be $\Gamma\sim$1-2.
More recently, the correlation has been extended to BHBs in quiescence, notably down to very low fluxes for the first known system A~0620-00 \cite{Gallo2006}, see also \cite{Gallo2014}).
All stellar-mass black holes have likely similar masses (a few to several solar masses). After a correction for mass is added, it was found that the correlation can be extended to Active Galactic Nuclei (\cite{Merloni2003}, see also \cite{Plotkin2015}), spanning 15 orders of magnitude in X-ray flux \cite{Gallo2006}. 

However, as more multi-wavelength observations became available, the situation has become more complex. 
A group of sources are found to follow a different correlation, limited to high fluxes and below the main one (often referred to as the ``radio-quiet'' branch, see \cite{Gallo2012}). Figure \ref{fig:corbel} shows the clear presence of two branches: the radio-quiet one is steeper, fitted with $\alpha\sim$1 \cite{Gallo2012}. Recently, the statistical significance of the presence of two separate correlations has been questioned by \cite{Gallo2014}.
The system H 1743-322 was observed to move from one correlation to the other: during outburst decay it started on the radio-quiet branch, then as radio luminosity decreased X-ray luminosity stalled until the system reached the radio-loud branch \cite{Coriat2011}. This clearly indicates that the difference between the two branches cannot be associated to fundamental parameters of the different systems, which would not change during the outburst (see \cite{Soleri2011}). \cite{Coriat2011} explored different possibilities and concluded that the transition between the two branches could be due to a transition from a radiatively efficient to a radiatively inefficient accretion flow, which would make the ``radio-quiet'' branch an ``X-ray loud'' one (see \cite{Koerding2006}).
To complicate matters even further, \cite{Russell2015} found that during its 2011 outburst MAXI J1836-194 followed a path in the X-radio plane which was significantly steeper and connected the two existing branches. Although the correlation appears to be less {\it universal} than previously assumed, it is clear that its interpretation can give important insights on both accretion and ejection. For instance, the evaporation/condensation model by \cite{Meyer-Hofmeister2014} would explain the second, X-ray loud, branch as due to the presence of an inner optically thick accretion disc which would increase the soft photon input available to Comptonization. As this inner disc cannot work at low accretion rate, this interpretation also explains why no lower-branch points are observed at low X-ray fluxes and the switch of H 1743-322.
At or after the LHS-HIMS transition the correlation breaks down (see also below), as the energy spectrum becomes more and more dominated by an optically thick accretion disc and the radio emission also shows non-monotonic variations (see \cite{Fender2004}). Interestingly, neutron-star LMXBs also show a correlation between radio and X-ray flux, with three important differences: (1) all NS systems are more radio-quiet at a given X-ray luminosity; (2) the correlation is steeper and (3) the radio emission is not suppressed when the source transits to the soft states\cite{Migliari2006}. This also points to the presence of a radiatively inefficient regime of accretion for BHBs.

At shorter wavelengths, IR and optical, the situation is complicated by the fact that the accretion disc also contributes to the flux and the jet contribution has to be estimated. In the prototypical system GX 339-4, in the LHS a clear positive correlation between X-ray and IR flux was found, which terminated abruptly when the source entered the HIMS and the relative contribution between jet and disc are expected to change \cite{Homan2005}. Additional observations have shown that the correlation extends for four orders of magnitude and evidence for a break around $10^{-3}$L$_{Edd}$ has been found \cite{Coriat2009}. Recently, a complete analysis of data from 33 systems, both hosting black holes and neutron stars, has been published \cite{Russell2006}. For BHBs, they found a correlation with power-law index 0.6 extending from quiescence to bright LHS, similar to that observed in the radio.

%% file: wind_outflows.tex
\section{Winds and  outflows}\label{sec:4}

Over the last couple of decades we have witnessed the discovery of a multitude of highly-ionized absorbers in high-resolution X-ray spectra from both BH and NS X-ray binaries. The first detections were obtained thanks to ASCA on the BH binaries GRO J1655-40 (\cite{Ueda1998}) and GRS 1915+105 (\cite{Kotani2000a}). Narrow absorption lines in the spectra of these systems, identified as Fe XXV and Fe XXVI indicated the first of a myriad of discoveries of photo-ionised plasmas in LMXBs that followed the launch of X-ray observatories such as Chandra, XMM-Newton and Suzaku. 

After the first detections, it was soon clear that the photo-ionised absorbers in the form of winds or atmospheres could be ubiquitous to all X-ray binaries (e.g. \cite{Parmar2002}), but only recently it has become increasingly clear that the presence of these plasmas could be key to our understanding of these systems. It has been suggested that the amount of mass that leaves the system once the plasmas are outflowing can be of the order of or significantly higher than the mass accretion rate transferred through the accretion disc (e.g. \cite{Ponti2012}), with crucial consequences on the accretion-outflow equilibrium at play in these systems. 

Much of the recent work on the accretion/ejection connection at work during the outburst of black-hole x-ray transients has focused on the X-ray/radio correlations (see e.g. \cite{Fender2004} and \cite{Fender2009}). It is now established that transient sources emerge from quiescence entering the low-hard state (see \cite{Belloni2011}), where a steady radio jet is ubiquitously seen. Then they rise in luminosity until the hard-to-soft transition takes place. During this transition strong relativistic ejections are often seen - observed as bright radio flares - and connected to the disappearance of the steady radio jets. 

Strong disc winds have been predominantly observed in the thermal-dominated high/soft state of BH transients (and most recently in NS systems, see \cite{Ponti2014}). Conversely, the presence of winds has been excluded by observations in the low/hard state, where no signature of wind-like outflows has been found so far with high level of confidence. Therefore, it seems natural to conclude that in the soft state the radio jets and relativistic ejections seem to be replaced by highly ionized accretion disc winds, that may play a crucial role in the physics of accretion and ejection around compact object. For instance, the effects of the winds on the entire system could be key in the outburst evolution of transient systems, influencing or even triggering the return to the hard state.

\subsection{Accretion disc winds and atmospheres}

To date, several LMXBs, both containing BHs and NSs (see e.g. \cite{DiazTrigo2006} and \cite{Ponti2012}), have shown narrow absorption lines, more often than not blue-shifted (i.e. indicating outflowing material) that have been interpreted as a consequence of the presence of highly-ionised material local to the source, opposed to the absorption due to the interstellar medium. Most of these sources are observed at a relatively high inclination angle (60-70$\deg$), pointing to a distribution of the ionizing plasma close to the accretion disc, with an equatorial or flared geometry. 

While the relative depth of the absorption lines detected in the high-resolution spectra of LMXBs allows to determine the column density and the ionization state of the plasma responsible of the lines, their blue-shift with respect to their theoretical wavelength provides information on the relative velocity of the plasma. The ionization degree of the plasma is described through the ionization parameter, defined as $\xi = L/nr^2$, where $L$ is the luminosity of the ionizing source, $n$ is the electron density and $r$ is the distance between the ionizing source and the plasma. 

The column density of the ionized plasma detected in LMXBs ranges between 10$^{21}$ cm$^{-2}$ and 10$^{24}$ cm$^{-2}$ and there is no obvious difference between  the densities measured for BHs and NSs systems. The vast majority of the systems for which $\log(\xi)$ has been measured, show high ionization degrees, with $\log(\xi) > 3$.  However, MAXI J1305-70 has shown both high and low ionization degrees, and GX 339-4 (see \cite{Shidatsu2013}) has shown only a low ionisation degree plasma. Interestingly, among the sources known to show ionization plasmas, GX 339-4 is possibly  the one at the lowest inclination: this suggest the possibility that there could be a stratification of the ionized plasma as a function of inclination above the accretion disc. On the other hand, GX 339-4 is affected by relatively low interstellar absorption, thus the detection of low-ionization plasma could be affected by a significant observational bias. 

Even though the majority of the ionized plasmas detected so far in LMXBs is characterized by significant intrinsic velocities, for a few sources there is no evidence of an intrinsic plasma velocity. This difference justifies the classification of ionized plasmas in disc winds/outflows and disc atmospheres. When a given system shows absorption lines with significant blue-shift and/or a P-Cygni profile, the ionized plasma is classified as \textit{outflow}. If, instead, a given system does not show a significant blue-shift in the absorption line, then the ionized plasma takes the name of \textit{atmosphere}. Clearly, the most important difference between the systems shown winds/outflows and the ones showing atmospheres is that only the former loose mass, with a rate that can be even significantly larger than the mass accretion rate (see e.g. \cite{Lee2002}, \cite{Neilsen2012}, \cite{Ponti2012}). 

Among the sources for which ionized plasmas has been detected, all BH LMXBs shows relatively fast outflows, while only about $\sim$30\% of NS LMXBs have shown clear winds (see \cite{Ponti2014}). This difference between NSs and BHS could be due to a significant difference in size of the system, i.e. strong winds are more easily produced in long orbital period (i.e. large systems, and, consequently, large discs). This hypothesis is supported by the fact that the NS systems producing winds are those with orbital periods comparable to the BH ones. 

\subsection{Winds launching mechanism}

Three main mechanisms have been identified as possible responsibles for the launching of fast winds from the accretion disc of accreting sources: thermal pressure, radiative pressure and magnetic pressure. The mechanism normally invoked to explain the strong winds detected in the radio-free soft states of BH and NS X-ray binaries is the thermal pressure, even though it is importan to bear in mind that the dominant wind launching mechanism could change as the system evolves.

\subsubsection{Thermally driven winds}

Thermal pressure or Compton-heated winds should arise in systems where the accretion disc is irradiated by the central regions of the accretion flow, such as in X-ray binaries or in quasars. The disc gas can be heated to temperatures exceeding 10$^7$ K mostly through the Compton process, partially evaporating and forming a corona above the disc. This gas, depending on the thermal velocity exceeding or not the local escape velocity, can be either emitted as a thermal wind or stay bound to the disc, forming an atmosphere (\cite{Begelman1983}). The radial extent of this corona only depends on the mass of the central compact object and on the Compton temperature, while it is independent on the luminosity. 
It has been shown (\cite{Begelman1983}) that due to disc rotation a wind can be launched via thermal pressure at radii larger than $\sim$ 0.1 $r_{C}$, where $r_{C}$ is the Compton radius, i.e. the radius where the escape velocity equals the isothermal sound speed at the Compton temperature $T_{C}$. For very luminous systems (where the radiation force due to electrons must be taken in account), \cite{Proga2002} found that strong winds can be launched starting from radii as small as 0.01 $r_{IC}$.

\subsubsection{Radiatively driven winds}

Radiation-driven winds might arise from an accretion disc when radiative acceleration occurs due to the transfer of momentum from the photon field to the corona, that leaves the disc in the form of an outflow. 
Assuming that the wind has optical depth $\tau_e$ higher than 1 and completely surrounds a source of radiation with luminosity L$_{bol}$, the multiple scattering of photons within the wind will lead to a wind momentum that exceed the photon momentum of the primary emission. In most sources, however, the material constituting the wind appears to be far too ionized to allow an effective momentum transfer and a consequent launching mechanism. In other words, the momentum-flux in radiatively-driven winds cannot exceed that of the radiation field, even considering the effects of radiation of free electrons (see \cite{Reynolds2012}).

\subsubsection{Magnetically driven winds}

disc winds can also be driven by magnetic forces. These winds are expected to be centrifugally accelerated down open, rotating, poloidal magnetic fields anchored in the disc (see \cite{Blandford1982}). Being these winds accelerated by the effect of magnetic torques from magnetic fields embedded in the accretion disc, there must be an intimate connection between the mass-loss in the wind and the accretion onto the black hole. According to the theory of MHD winds, material is centrifugally lifted off the disc at a certain launching radius R$_L$ and is continuously accelerated by magnetic forces until the flow become super-Alfvenic at a radius R$_A >$ R$_L$. Accurate models suggest that the ratio R$_A$/R$_L$ ranges between 2 and 3 (see \cite{Pudritz2007}). \cite{Reynolds2012} has shown that MHD torques would be able to produce Compton-thick winds only if  (1) the accretion rate is a significant fraction of the Eddington rate, (2) the radiative efficiency is low and (3) the Alfven radius is very close to the launching radius (R$_A$ $sim$R$_L$). Thus, MHD driving could become a viable explanation for disc winds only in a very limited number of cases.

%% file: the_full_picture.tex
\section{The full accretion ejection picture}\label{subsubsec:3.2.1}

Despite the number of complications and additional details, the overall picture is now much more clear than before the RXTE data. Not all transient sources behave in the fundamental diagrams as neatly as the one shown in Fig. 1, but the general pattern appears to be followed by most sources and this regularity points towards fundamental aspects of accretion onto black holes (see \cite{Fender2009}). Indeed, a comparison with different classes of related sources shows that this must be the case.

{\it Persistent BHBs.} A few persistent systems are known in our galaxy and the Magellanic Clouds. Some of them appear locked in a single state. LMC X-1 has always been observed in the HSS (see \cite{Ruhlen2011}), as the bright galactic source 4U~1957+11 \cite{Nowak2008}. The second BHB in the LMC, LMC X-3, was only observed in the HSS until with the extensive coverage of RXTE brief transitions to and from the LHS were discovered \cite{Smale2012}. The source GRS 1758-258 in the Galactic Center region is mostly in the LHS, but shows sporadic transitions to the HSS (see \cite{Soria2011}). The bright source 4U 1755-33 was very bright and in the HSS until 1996, when it went into quiescence (see \cite{Angelini2003}), but unfortunately the decay into quiescence was not covered by observations. The brightest and best known Cyg X-1, the first black-hole candidate, is usually found in the LHS and makes rather frequent transitions to the HIMS, to rarely reach the HSS when radio emission is observed to drop (see \cite{Grinberg2013} and references therein. A HID from the RXTE observations of Cyg X-1 is shown in \cite{Belloni2010}. Overall, none of these systems shows a behavior inconsistent with the above picture, although clearly none of them shows a full ``transient'' cycle. In particular, the LHS-HIMS and HMIS-LHS transitions of Cyg X-1 do not show any sign of hysteresis. Interestingly, corresponding to one of the softest events of Cyg X-1, a radio flare was observed, compatible with a jet ejection in correspondence to a transition to the HSS \cite{Wilms2007}.

A case of its own is represented by GRS 1915+105, which has started an outburst in 1992 and is still active at the time of writing. Its behavior is very different from all other systems, although a few sources have been found to match it rather precisely for some limited time \cite{Altamirano2011,Bagnoli2015}. The original idea was that this peculiar behavior is connected to a very high accretion rate, which would put the source above the standard `q' diagram (see \cite{Belloni2010} for an HID of state-C, i.e. hard, intervals of GRS 1915+105). However, recently the same type of structured variability has been found in the Rapid Burster, a neutron-star binary, at luminosities well below the Eddington limit, casting doubts on this interpretation \cite{Bagnoli2015}. At any rate, the short-term variability of GRS 1915+105 is not different from those observed in other BHBs: during state-C, whose energy spectrum corresponds to a LHS/HIMS, the PDS is a typical LHS/HIMS \cite{Reig2000,Reig2003,Belloni2010}, during softer and brighter states (called A and B, see \cite{Belloni2000} the PDS is similar to that of ``anomalous'' states at high accretion rate see in other BHBs \cite{Reig2003,Belloni2010}. HFQPOs are observed only in a very narrow region of the HID \cite{Belloni2013a}. Evidence for a type-B QPO during fast transition has also been presented \cite{soleri2008}. What is peculiar here is the structure of alternating states.

\begin{figure}[t]
	\includegraphics[scale=0.45,angle =270]{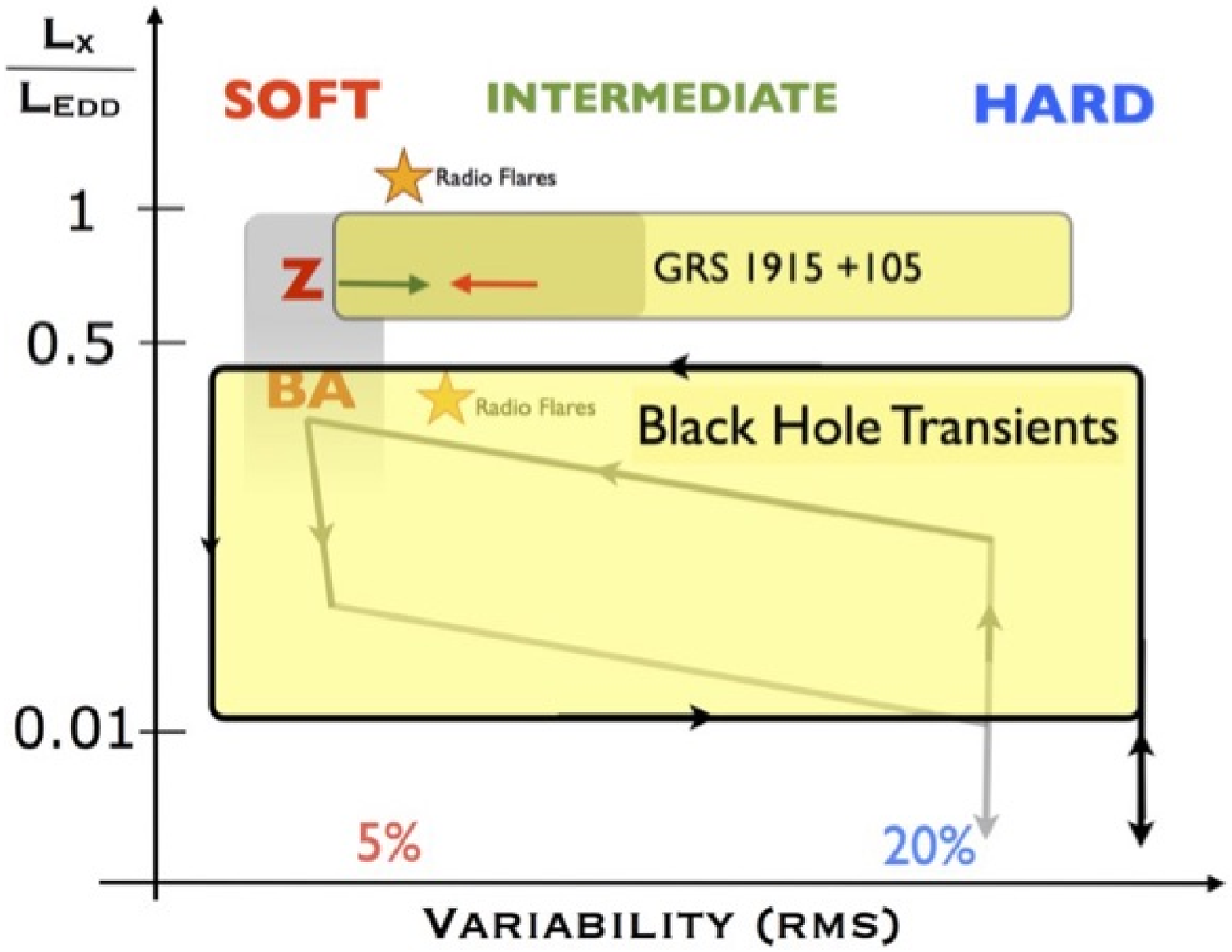}
	\caption{RID diagram with the regions occupied by BHBs and different classes of NS LMXBs (Z: Z sources, BA: bright atoll sources, the gray ``q'' corresponds to atoll sources; Fig. 10 from \cite{Munoz-Darias2014}). The X axis in this diagram is fractional rms variability, which correlates almost linearly with hardness, which means a HID would look almost identical.}
	\label{fig:teo}
\end{figure}

{\it Neutron-star binaries.} Neutron-star low-mass X-ray binaries show many similarities in their emission properties with BHBs. Their detailed X-ray energy spectra are rather different as the component of direct emission from the surface of the compact object and the boundary layer is very bright. However, it is now clear that the properties of fast time variability can be strongly connected to those of BHBs (see \cite{Olive1998,Casella2005,Wijnands1999a,Psaltis1999,Belloni2002}. The first evidence of a hysteresis pattern in the HID was presented from the low-luminosity persistent system 4U 1636-53 \cite{Belloni2007}. ``q'' diagrams were shown for the outbursts of Aql X-1 \cite{Belloni2010,Koerding2007}. More recently, a full analysis of RXTE data has shown that the observed pattern is very similar and a strong connection has been drawn \cite{Munoz-Darias2014}. As the difference between classes of neutron-star LMXBs has been finally attributed to accretion rate levels (see \cite{Homan2007}, a complete diagram for NS and BH binaries could be produced (see Fig. \ref{fig:teo}). A strong connection with the BHBs was found in the radio emission of Aql X-1, completing the picture \cite{Koerding2007}.
The radio-X correlation for NS binaries lies below that of BHBs, suggesting a difference in the emission efficiency between the two classes of systems in their hard state \cite{Migliari2006}.

{\it ULX.} In the case of Ultra-Luminous X-ray sources (ULX) the comparison is not simple, given the lower statistics and sparser coverage of observations, not to mention the fact that there is no agreement on the nature of the systems (see \cite{Feng2011} for a review). There are similarities in the timing properties, but there is no consensus on the interpretation of the few QPOs detected in ULX in terms of those in BHBs (see e.g. \cite{Strohmayer2003a,Mucciarelli2006,Pasham2013}). The energy spectra are even more complex and only basic comparisons can be made, not always in agreement with each other \cite{Belloni2011a,Soria2011a}. State transitions have been observed (see \cite{Kaaret2013} and references therein, but only with few observations. When standard diagrams have been produced for a sample of ULX, the results were also complex and difficult to interpret \cite{Pintore2014}.
In the case of the hyper luminous source ESO 243-49 HLX-1, a number of XMM-Newton and Chandra observations were used to produce HID and HRD, showing compatibility with the BHB diagrams \cite{Servillat2011}. Radio flares have been observed from this source, although the lack of high-SNR X-ray coverage prevents a precise association with source states \cite{Webb2012}.

{\it AGN} The association between AGN properties and BHB states is also complex. While Quasars and Blazars are thought to be jet-dominated systems and Seyfert II AGN suffer of large absorption, a comparison can be attempted with Seyfert I systems. On the spectral side, the energy distribution in X-rays is similar to that of galactic systems in the LHS, although for AGN the thermal accretion disc component is not in the ``standard'' X-ray band. Indeed, the production of a HID from RXTE data of Sy I objects shows that almost all of them lie on the LHS branch. However, analysis of variability, which in the case of AGN must be observed up to time scales of years, shows that their PDS is more similar to that of HSS binaries (see .e.g. \cite{McHardy2005,McHardy2010}). Two systems stand out both spectrally and in timing properties: Ton S180 and Ark 564. When placed on a standard HID, renormalized for flux differences, these object lie in an intermediate-hardness area \cite{Belloni2010}. The exact connection to the BHB diagram is not easy as the horizontal branches are caused by both a softening of the LHS component and the appearance of a thermal disc component, absent in the AGN 2-10 keV spectra, but the position is certainly intermediate. On the timing side, these two objects also show properties which can be identified of more intermediate states \cite{Arevalo2006,McHardy2007,McHardy2010}.

This general behavior must be understood in terms of basic parameters. Interpretations of the overall evolution in the HID have been proposed \cite{Maccarone2003,Begelman2014,Begelman2015}. However, the correlated properties of the emission of relativistic jets and winds from the system must be included in order to reach a global understanding. This of course complicates the problem and only basic attempts have been made \cite{Fender2012,Fender2004}. A zero-level approach has been proposed recently, which in addition to the X-ray properties incorporates a model for the production of poloidal magnetic fields, which can be a crucial ingredient for the production of jets \cite{Kylafis2015}. The ``cosmic-battery'' scenario  (see \cite{Contopoulos1998,Kylafis2012,Contopoulos2012,Contopoulos2015}) can be the link between properties of the accretion flow and generation of relativistic jets, which is worth exploring further.

%% file: conclusions.tex

Thanks to the availability of high-quality data from past and current high-energy missions, our knowledge and understanding of black-hole transients has increased significantly in the past two decades. Instrumental to this advancements have been observations at other wavelengths, which have allowed us to study physical components, such as relativistic jets and outflows, that were completely ignored before. It is now clear that studying the emission over a very broad range of energies is the only way to properly characterize the physical properties of these objects. For the study of fast X-ray time variability the gap left by the demise of RossiXTE has just been filled by the launch of the indian satellite ASTROSAT, which at the time of writing is still in the performance-verification phase. 
Theoretical modelling of observational data is now turning to the interpretation of joint spectral-timing properties, a crucial step to study rapidly varying phenomena. The field is evolving very rapidly: however, while the amount of available details make the full picture more complicated, the basic patterns are converging towards a solid set of observational points that must be at the base of all theoretical model.
Above, we have focussed onto those patterns with the aim of outlining general properties of black-hole transients. This is by no means complete (see Middleton's chapter on detailed spectral models), but it should constitute a solid starting point upon which layers of complexity can later be laid.

%% file: Belloni_Motta.bbl
\begin{thebibliography}{100}

\bibitem{Abramowicz2001}
M.~A. {Abramowicz} and W.~{Klu{\'z}niak}.
\newblock {A precise determination of black hole spin in GRO J1655-40}.
\newblock {\em \aap}, 374:L19--L20, August 2001.

\bibitem{Abramowicz2004}
M.~A. {Abramowicz}, W.~{Kluzniak}, Z.~{Stuchlik}, and G.~{Torok}.
\newblock {The orbital resonance model for twin peak kHz QPOs}.
\newblock {\em ArXiv Astrophysics e-prints}, January 2004.

\bibitem{Altamirano2012}
D.~{Altamirano} and T.~{Belloni}.
\newblock {Discovery of High-frequency Quasi-periodic Oscillations in the Black
  Hole Candidate IGR J17091-3624}.
\newblock {\em \apjl}, 747:L4, March 2012.

\bibitem{Altamirano2011}
D.~{Altamirano}, T.~{Belloni}, M.~{Linares}, M.~{van der Klis}, R.~{Wijnands},
  P.~A. {Curran}, M.~{Kalamkar}, H.~{Stiele}, S.~{Motta},
  T.~{Mu{\~n}oz-Darias}, P.~{Casella}, and H.~{Krimm}.
\newblock {The Faint ''Heartbeats'' of IGR J17091-3624: An Exceptional Black
  Hole Candidate}.
\newblock {\em \apjl}, 742:L17, December 2011.

\bibitem{Angelini2003}
L.~{Angelini} and N.~E. {White}.
\newblock {An XMM-Newton Observation of 4U 1755-33 in Quiescence: Evidence of a
  Fossil X-Ray Jet}.
\newblock {\em \apjl}, 586:L71--L75, March 2003.

\bibitem{Arevalo2006}
P.~{Ar{\'e}valo}, I.~E. {Papadakis}, P.~{Uttley}, I.~M. {McHardy}, and
  W.~{Brinkmann}.
\newblock {Spectral-timing evidence for a very high state in the narrow-line
  Seyfert 1 Ark 564}.
\newblock {\em \mnras}, 372:401--409, October 2006.

\bibitem{ArmasPadilla2011}
M.~{Armas Padilla}, N.~{Degenaar}, A.~{Patruno}, D.~M. {Russell}, M.~{Linares},
  T.~J. {Maccarone}, J.~{Homan}, and R.~{Wijnands}.
\newblock {X-ray softening in the new X-ray transient XTE J1719-291 during its
  2008 outburst decay}.
\newblock {\em \mnras}, 417:659--665, October 2011.

\bibitem{Bachetti2014}
M.~{Bachetti}, F.~A. {Harrison}, D.~J. {Walton}, B.~W. {Grefenstette},
  D.~{Chakrabarty}, F.~{F{\"u}rst}, D.~{Barret}, A.~{Beloborodov}, S.~E.
  {Boggs}, F.~E. {Christensen}, W.~W. {Craig}, A.~C. {Fabian}, C.~J. {Hailey},
  A.~{Hornschemeier}, V.~{Kaspi}, S.~R. {Kulkarni}, T.~{Maccarone}, J.~M.
  {Miller}, V.~{Rana}, D.~{Stern}, S.~P. {Tendulkar}, J.~{Tomsick}, N.~A.
  {Webb}, and W.~W. {Zhang}.
\newblock {An ultraluminous X-ray source powered by an accreting neutron star}.
\newblock {\em \nat}, 514:202--204, October 2014.

\bibitem{Bagnoli2015}
T.~{Bagnoli} and J.~J.~M. {in't Zand}.
\newblock {Discovery of GRS 1915+105 variability patterns in the Rapid
  Burster}.
\newblock {\em \mnras}, 450:L52--L56, June 2015.

\bibitem{Begelman2014}
M.~C. {Begelman} and P.~J. {Armitage}.
\newblock {A Mechanism for Hysteresis in Black Hole Binary State Transitions}.
\newblock {\em \apjl}, 782:L18, February 2014.

\bibitem{Begelman2015}
M.~C. {Begelman}, P.~J. {Armitage}, and C.~S. {Reynolds}.
\newblock {Accretion Disk Dynamo as the Trigger for X-Ray Binary State
  Transitions}.
\newblock {\em \apj}, 809:118, August 2015.

\bibitem{Begelman1983}
M.~C. {Begelman}, C.~F. {McKee}, and G.~A. {Shields}.
\newblock {Compton heated winds and coronae above accretion disks. I Dynamics}.
\newblock {\em \apj}, 271:70--88, August 1983.

\bibitem{Belloni2005a}
T.~{Belloni}, J.~{Homan}, P.~{Casella}, M.~{van der Klis}, E.~{Nespoli},
  W.~H.~G. {Lewin}, J.~M. {Miller}, and M.~{M{\'e}ndez}.
\newblock {The evolution of the timing properties of the black-hole transient
  GX 339-4 during its 2002/2003 outburst}.
\newblock {\em \aap}, 440:207--222, September 2005.

\bibitem{Belloni2007}
T.~{Belloni}, J.~{Homan}, S.~{Motta}, E.~{Ratti}, and M.~{M{\'e}ndez}.
\newblock {Rossi XTE monitoring of 4U1636-53 - I. Long-term evolution and kHz
  quasi-periodic oscillations}.
\newblock {\em \mnras}, 379:247--252, July 2007.

\bibitem{Belloni2000}
T.~{Belloni}, M.~{Klein-Wolt}, M.~{M{\'e}ndez}, M.~{van der Klis}, and J.~{van
  Paradijs}.
\newblock {A model-independent analysis of the variability of GRS 1915+105}.
\newblock {\em \aap}, 355:271--290, March 2000.

\bibitem{Belloni2001}
T.~{Belloni}, M.~{M{\'e}ndez}, and C.~{S{\'a}nchez-Fern{\'a}ndez}.
\newblock {The high-frequency QPOs in GRS 1915+105}.
\newblock {\em \aap}, 372:551--556, June 2001.

\bibitem{Belloni1996}
T.~{Belloni}, M.~{Mendez}, M.~{van der Klis}, G.~{Hasinger}, W.~H.~G. {Lewin},
  and J.~{van Paradijs}.
\newblock {An Intermediate State of Cygnus X-1}.
\newblock {\em \apjl}, 472:L107, December 1996.

\bibitem{Belloni2002}
T.~{Belloni}, D.~{Psaltis}, and M.~{van der Klis}.
\newblock {A Unified Description of the Timing Features of Accreting X-Ray
  Binaries}.
\newblock {\em \apj}, 572:392--406, June 2002.

\bibitem{Belloni1997}
T.~{Belloni}, M.~{van der Klis}, W.~H.~G. {Lewin}, J.~{van Paradijs},
  T.~{Dotani}, K.~{Mitsuda}, and S.~{Miyamoto}.
\newblock {Energy dependence in the quasi-periodic oscillations and noise of
  black hole candidates in the very high state.}
\newblock {\em \aap}, 322:857--867, June 1997.

\bibitem{Belloni2010}
T.~M. {Belloni}.
\newblock {States and Transitions in Black Hole Binaries}.
\newblock {\em Lecture Notes in Physics, Springer-Verlag Berlin Heidelberg,
  Volume 794, p. 53. ISBN 978-3-540-76936-1.}, 794:53--+, March 2010.

\bibitem{Belloni2011a}
T.~M. {Belloni}.
\newblock {Black-hole states in external galaxies}.
\newblock {\em Astronomische Nachrichten}, 332:324, May 2011.

\bibitem{Belloni2013}
T.~M. {Belloni} and D.~{Altamirano}.
\newblock {Discovery of a 34 Hz quasi-periodic oscillation in the X-ray
  emission of GRS 1915+105}.
\newblock {\em \mnras}, 432:19--22, June 2013.

\bibitem{Belloni2013a}
T.~M. {Belloni} and D.~{Altamirano}.
\newblock {High-frequency quasi-periodic oscillations from GRS 1915+105}.
\newblock {\em \mnras}, 432:10--18, June 2013.

\bibitem{Belloni2011}
T.~M. {Belloni}, S.~E. {Motta}, and T.~{Mu{\~n}oz-Darias}.
\newblock {Black hole transients}.
\newblock {\em Bulletin of the Astronomical Society of India}, 39:409--428,
  September 2011.

\bibitem{Belloni2012}
T.~M. {Belloni}, A.~{Sanna}, and M.~{M{\'e}ndez}.
\newblock {High-frequency quasi-periodic oscillations in black hole binaries}.
\newblock {\em \mnras}, 426:1701--1709, November 2012.

\bibitem{Belloni2014}
T.~M. {Belloni} and L.~{Stella}.
\newblock {Fast Variability from Black-Hole Binaries}.
\newblock {\em \ssr}, 183:43--60, September 2014.

\bibitem{Blandford1982}
R.~D. {Blandford} and D.~G. {Payne}.
\newblock {Hydromagnetic flows from accretion discs and the production of radio
  jets}.
\newblock {\em \mnras}, 199:883--903, June 1982.

\bibitem{Brocksopp2004}
C.~{Brocksopp}, R.~M. {Bandyopadhyay}, and R.~P. {Fender}.
\newblock {``Soft X-ray transient'' outbursts which are not soft}.
\newblock {\em New Astronomy}, 9:249--264, May 2004.

\bibitem{Brocksopp2002}
C.~{Brocksopp}, R.~P. {Fender}, M.~{McCollough}, G.~G. {Pooley}, M.~P. {Rupen},
  R.~M. {Hjellming}, C.~J. {de la Force}, R.~E. {Spencer}, T.~W.~B. {Muxlow},
  S.~T. {Garrington}, and S.~{Trushkin}.
\newblock {Initial low/hard state, multiple jet ejections and X-ray/radio
  correlations during the outburst of XTE J1859+226}.
\newblock {\em \mnras}, 331:765--775, April 2002.

\bibitem{Brocksopp2010}
C.~{Brocksopp}, P.~G. {Jonker}, D.~{Maitra}, H.~A. {Krimm}, G.~G. {Pooley},
  G.~{Ramsay}, and C.~{Zurita}.
\newblock {Disentangling jet and disc emission from the 2005 outburst of XTE
  J1118+480}.
\newblock {\em \mnras}, 404:908--916, May 2010.

\bibitem{Bursa2005}
M.~{Bursa}.
\newblock {High-frequency QPOs in GRO J1655-40: Constraints on resonance models
  by spectral fits}.
\newblock In S.~{Hled{\'{\i}}k} and Z.~{Stuchl{\'{\i}}k}, editors, {\em RAGtime
  6/7: Workshops on black holes and neutron stars}, pages 39--45, December
  2005.

\bibitem{Cabanac2010}
C.~{Cabanac}, G.~{Henri}, P.-O. {Petrucci}, J.~{Malzac}, J.~{Ferreira}, and
  T.~M. {Belloni}.
\newblock {Variability of X-ray binaries from an oscillating hot corona}.
\newblock {\em \mnras}, 404:738--748, May 2010.

\bibitem{Capitanio2009}
F.~{Capitanio}, T.~{Belloni}, M.~{Del Santo}, and P.~{Ubertini}.
\newblock {A failed outburst of H1743-322}.
\newblock {\em \mnras}, 398:1194--1200, September 2009.

\bibitem{Casella2004}
P.~{Casella}, T.~{Belloni}, J.~{Homan}, and L.~{Stella}.
\newblock {A study of the low-frequency quasi-periodic oscillations in the
  X-ray light curves of the black hole candidate <ASTROBJ>XTE
  J1859+226</ASTROBJ>}.
\newblock {\em \aap}, 426:587--600, November 2004.

\bibitem{Casella2005}
P.~{Casella}, T.~{Belloni}, and L.~{Stella}.
\newblock {The ABC of Low-Frequency Quasi-periodic Oscillations in Black Hole
  Candidates: Analogies with Z Sources}.
\newblock {\em \apj}, 629:403--407, August 2005.

\bibitem{Contopoulos1998}
I.~{Contopoulos} and D.~{Kazanas}.
\newblock {A Cosmic Battery}.
\newblock {\em \apj}, 508:859--863, December 1998.

\bibitem{Contopoulos2015}
I.~{Contopoulos}, A.~{Nathanail}, and M.~{Katsanikas}.
\newblock {The Cosmic Battery in Astrophysical Accretion Disks}.
\newblock {\em \apj}, 805:105, June 2015.

\bibitem{Contopoulos2012}
I.~{Contopoulos} and D.~B. {Papadopoulos}.
\newblock {The Cosmic Battery and the inner edge of the accretion disc}.
\newblock {\em \mnras}, 425:147--152, September 2012.

\bibitem{Corbel2013}
S.~{Corbel}, M.~{Coriat}, C.~{Brocksopp}, A.~K. {Tzioumis}, R.~P. {Fender},
  J.~A. {Tomsick}, M.~M. {Buxton}, and C.~D. {Bailyn}.
\newblock {The `universal' radio/X-ray flux correlation: the case study of the
  black hole GX 339-4}.
\newblock {\em \mnras}, 428:2500--2515, January 2013.

\bibitem{Corbel2003}
S.~{Corbel}, M.~A. {Nowak}, R.~P. {Fender}, A.~K. {Tzioumis}, and S.~{Markoff}.
\newblock {Radio/X-ray correlation in the low/hard state of GX 339-4}.
\newblock {\em \aap}, 400:1007--1012, March 2003.

\bibitem{Coriat2009}
M.~{Coriat}, S.~{Corbel}, M.~M. {Buxton}, C.~D. {Bailyn}, J.~A. {Tomsick},
  E.~{K{\"o}rding}, and E.~{Kalemci}.
\newblock {The infrared/X-ray correlation of GX 339-4: probing hard X-ray
  emission in accreting black holes}.
\newblock {\em \mnras}, 400:123--133, November 2009.

\bibitem{Coriat2011}
M.~{Coriat}, S.~{Corbel}, L.~{Prat}, J.~C.~A. {Miller-Jones}, D.~{Cseh}, A.~K.
  {Tzioumis}, C.~{Brocksopp}, J.~{Rodriguez}, R.~P. {Fender}, and G.~R.
  {Sivakoff}.
\newblock {Radiatively efficient accreting black holes in the hard state: the
  case study of H1743-322}.
\newblock {\em \mnras}, 414:677--690, June 2011.

\bibitem{DiazTrigo2006}
M.~{D{\'{\i}}az Trigo}, A.~N. {Parmar}, L.~{Boirin}, M.~{M{\'e}ndez}, and J.~S.
  {Kaastra}.
\newblock {Spectral changes during dipping in low-mass X-ray binaries due to
  highly-ionized absorbers}.
\newblock {\em \aap}, 445:179--195, January 2006.

\bibitem{Done2007}
C.~{Done}, M.~{Gierlinski}, and A.~{Kubota}.
\newblock {Modelling the behaviour of accretion flows in X-ray binaries.
  Everything you always wanted to know about accretion but were afraid to ask}.
\newblock {\em \aap}, 15:1--66, December 2007.

\bibitem{Done2006}
C.~{Done} and A.~{Kubota}.
\newblock {Disc-corona energetics in the very high state of Galactic black
  holes}.
\newblock {\em \mnras}, 371:1216--1230, September 2006.

\bibitem{Dunn2010}
R.~J.~H. {Dunn}, R.~P. {Fender}, E.~G. {K{\"o}rding}, T.~{Belloni}, and
  C.~{Cabanac}.
\newblock {A global spectral study of black hole X-ray binaries}.
\newblock {\em \mnras}, 403:61--82, March 2010.

\bibitem{Elvis1975}
M.~{Elvis}, C.~G. {Page}, K.~A. {Pounds}, M.~J. {Ricketts}, and M.~J.~L.
  {Turner}.
\newblock {Discovery of powerful transient X-ray source A0620-00 with Ariel V
  Sky Survey Experiment}.
\newblock {\em \nat}, 257:656--+, October 1975.

\bibitem{Esin1997}
A.~A. {Esin}, J.~E. {McClintock}, and R.~{Narayan}.
\newblock {Advection-dominated Accretion and the Spectral States of Black Hole
  X-Ray Binaries: Application to Nova MUSCAE 1991}.
\newblock {\em \apj}, 489:865--+, November 1997.

\bibitem{Fender2012}
R.~{Fender} and T.~{Belloni}.
\newblock {Stellar-Mass Black Holes and Ultraluminous X-ray Sources}.
\newblock {\em Science}, 337:540--, August 2012.

\bibitem{Fender2014}
R.~{Fender} and E.~{Gallo}.
\newblock {An Overview of Jets and Outflows in Stellar Mass Black Holes}.
\newblock {\em \ssr}, 183:323--337, September 2014.

\bibitem{Fender2004}
R.~P. {Fender}, T.~M. {Belloni}, and E.~{Gallo}.
\newblock {Towards a unified model for black hole X-ray binary jets}.
\newblock {\em \mnras}, 355:1105--1118, December 2004.

\bibitem{Fender2009}
R.~P. {Fender}, J.~{Homan}, and T.~M. {Belloni}.
\newblock {Jets from black hole X-ray binaries: testing, refining and extending
  empirical models for the coupling to X-rays}.
\newblock {\em \mnras}, 396:1370--1382, July 2009.

\bibitem{Feng2011}
H.~{Feng} and R.~{Soria}.
\newblock {Ultraluminous X-ray sources in the Chandra and XMM-Newton era}.
\newblock {\em \nar}, 55:166--183, November 2011.

\bibitem{Ferreira2006}
J.~{Ferreira}, P.-O. {Petrucci}, G.~{Henri}, L.~{Saug{\'e}}, and
  G.~{Pelletier}.
\newblock {A unified accretion-ejection paradigm for black hole X-ray binaries.
  I. The dynamical constituents}.
\newblock {\em \aap}, 447:813--825, March 2006.

\bibitem{Gallo2010}
E.~{Gallo}.
\newblock {Radio Emission and Jets from Microquasars}.
\newblock In {T.~Belloni}, editor, {\em Lecture Notes in Physics, Berlin
  Springer Verlag}, volume 794 of {\em Lecture Notes in Physics, Berlin
  Springer Verlag}, pages 85--+, March 2010.

\bibitem{Gallo2006}
E.~{Gallo}, R.~P. {Fender}, J.~C.~A. {Miller-Jones}, A.~{Merloni}, P.~G.
  {Jonker}, S.~{Heinz}, T.~J. {Maccarone}, and M.~{van der Klis}.
\newblock {A radio-emitting outflow in the quiescent state of A0620-00:
  implications for modelling low-luminosity black hole binaries}.
\newblock {\em \mnras}, 370:1351--1360, August 2006.

\bibitem{Gallo2003}
E.~{Gallo}, R.~P. {Fender}, and G.~G. {Pooley}.
\newblock {A universal radio-X-ray correlation in low/hard state black hole
  binaries}.
\newblock {\em \mnras}, 344:60--72, September 2003.

\bibitem{Gallo2012}
E.~{Gallo}, B.~P. {Miller}, and R.~{Fender}.
\newblock {Assessing luminosity correlations via cluster analysis: evidence for
  dual tracks in the radio/X-ray domain of black hole X-ray binaries}.
\newblock {\em \mnras}, 423:590--599, June 2012.

\bibitem{Gallo2014}
E.~{Gallo}, J.~C.~A. {Miller-Jones}, D.~M. {Russell}, P.~G. {Jonker},
  J.~{Homan}, R.~M. {Plotkin}, S.~{Markoff}, B.~P. {Miller}, S.~{Corbel}, and
  R.~P. {Fender}.
\newblock {The radio/X-ray domain of black hole X-ray binaries at the lowest
  radio luminosities}.
\newblock {\em \mnras}, 445:290--300, November 2014.

\bibitem{Gierlinski2008}
M.~{Gierli{\'n}ski}, M.~{Middleton}, M.~{Ward}, and C.~{Done}.
\newblock {A periodicity of \~{}1hour in X-ray emission from the active galaxy
  RE J1034+396}.
\newblock {\em \nat}, 455:369--371, September 2008.

\bibitem{Gierlinski1999}
M.~{Gierli{\'n}ski}, A.~A. {Zdziarski}, J.~{Poutanen}, P.~S. {Coppi},
  K.~{Ebisawa}, and W.~N. {Johnson}.
\newblock {Radiation mechanisms and geometry of Cygnus X-1 in the soft state}.
\newblock {\em \mnras}, 309:496--512, October 1999.

\bibitem{Gilfanov2010}
M.~{Gilfanov}.
\newblock {X-Ray Emission from Black-Hole Binaries}.
\newblock {\em The Jet Paradigm, Lecture Notes in Physics, Springer-Verlag
  Berlin Heidelberg, Volume 794, p. 17. ISBN 978-3-540-76936-1.}, 794:17--+,
  March 2010.

\bibitem{Grinberg2013}
V.~{Grinberg}, N.~{Hell}, K.~{Pottschmidt}, M.~{B{\"o}ck}, M.~A. {Nowak},
  J.~{Rodriguez}, A.~{Bodaghee}, M.~{Cadolle Bel}, G.~L. {Case}, M.~{Hanke},
  M.~{K{\"u}hnel}, S.~B. {Markoff}, G.~G. {Pooley}, R.~E. {Rothschild}, J.~A.
  {Tomsick}, C.~A. {Wilson-Hodge}, and J.~{Wilms}.
\newblock {Long term variability of Cygnus X-1. V. State definitions with all
  sky monitors}.
\newblock {\em \aap}, 554:A88, June 2013.

\bibitem{Haardt1993}
F.~{Haardt} and L.~{Maraschi}.
\newblock {X-ray spectra from two-phase accretion disks}.
\newblock {\em \apj}, 413:507--517, August 1993.

\bibitem{Homan2005a}
J.~{Homan} and T.~{Belloni}.
\newblock {The Evolution of Black Hole States}.
\newblock {\em \apss}, 300:107--117, November 2005.

\bibitem{Homan2005}
J.~{Homan}, M.~{Buxton}, S.~{Markoff}, C.~D. {Bailyn}, E.~{Nespoli}, and
  T.~{Belloni}.
\newblock {Multiwavelength Observations of the 2002 Outburst of GX 339-4: Two
  Patterns of X-Ray-Optical/Near-Infrared Behavior}.
\newblock {\em \apj}, 624:295--306, May 2005.

\bibitem{Homan2002a}
J.~{Homan}, M.~{van der Klis}, P.~G. {Jonker}, R.~{Wijnands}, E.~{Kuulkers},
  M.~{M{\'e}ndez}, and W.~H.~G. {Lewin}.
\newblock {RXTE Observations of the Neutron Star Low-Mass X-Ray Binary GX 17+2:
  Correlated X-Ray Spectral and Timing Behavior}.
\newblock {\em \apj}, 568:878--900, April 2002.

\bibitem{Homan2007}
J.~{Homan}, M.~{van der Klis}, R.~{Wijnands}, T.~{Belloni}, R.~{Fender},
  M.~{Klein-Wolt}, P.~{Casella}, M.~{M{\'e}ndez}, E.~{Gallo}, W.~H.~G. {Lewin},
  and N.~{Gehrels}.
\newblock {Rossi X-Ray Timing Explorer Observations of the First Transient Z
  Source XTE J1701-462: Shedding New Light on Mass Accretion in Luminous
  Neutron Star X-Ray Binaries}.
\newblock {\em \apj}, 656:420--430, February 2007.

\bibitem{Homan2001}
J.~{Homan}, R.~{Wijnands}, M.~{van der Klis}, T.~{Belloni}, J.~{van Paradijs},
  M.~{Klein-Wolt}, R.~{Fender}, and M.~{M{\'e}ndez}.
\newblock {Correlated X-Ray Spectral and Timing Behavior of the Black Hole
  Candidate XTE J1550-564: A New Interpretation of Black Hole States}.
\newblock {\em \apjs}, 132:377--402, February 2001.

\bibitem{Ingram2009}
A.~{Ingram}, C.~{Done}, and P.~C. {Fragile}.
\newblock {Low-frequency quasi-periodic oscillations spectra and Lense-Thirring
  precession}.
\newblock {\em \mnras}, 397:L101--L105, July 2009.

\bibitem{Ingram2013}
A.~{Ingram} and M.~v.~d. {van der Klis}.
\newblock {An exact analytic treatment of propagating mass accretion rate
  fluctuations in X-ray binaries}.
\newblock {\em \mnras}, 434:1476--1485, September 2013.

\bibitem{Kaaret2013}
P.~{Kaaret} and H.~{Feng}.
\newblock {A State Transition of the Luminous X-Ray Binary in the
  Low-metallicity Blue Compact Dwarf Galaxy I Zw 18}.
\newblock {\em \apj}, 770:20, June 2013.

\bibitem{Kalemci2014}
E.~{Kalemci}, M.~{\"O}. {Arabac{\i}}, T.~{G{\"u}ver}, D.~M. {Russell}, J.~A.
  {Tomsick}, J.~{Wilms}, G.~{Weidenspointner}, E.~{Kuulkers}, M.~{Falanga},
  T.~{Din{\c c}er}, S.~{Drave}, T.~{Belloni}, M.~{Coriat}, F.~{Lewis}, and
  T.~{Mu{\~n}oz-Darias}.
\newblock {Multiwavelength observations of the black hole transient Swift
  J1745-26 during the outburst decay}.
\newblock {\em \mnras}, 445:1288--1298, December 2014.

\bibitem{Kalemci2013}
E.~{Kalemci}, T.~{Din{\c c}er}, J.~A. {Tomsick}, M.~M. {Buxton}, C.~D.
  {Bailyn}, and Y.~Y. {Chun}.
\newblock {Complete Multiwavelength Evolution of Galactic Black Hole Transients
  during Outburst Decay. I. Conditions for ''Compact'' Jet Formation}.
\newblock {\em \apj}, 779:95, December 2013.

\bibitem{Kato2004}
S.~{Kato}.
\newblock {Resonant Excitation of Disk Oscillations by Warps: A Model of kHz
  QPOs}.
\newblock {\em \pasj}, 56:905--922, October 2004.

\bibitem{Kato2004a}
S.~{Kato}.
\newblock {Wave-Warp Resonant Interactions in Relativistic Disks and kHz QPOs}.
\newblock {\em \pasj}, 56:559--567, June 2004.

\bibitem{Kato2005}
S.~{Kato}.
\newblock {A Resonance Model of Quasi-Periodic Oscillations of Low-Mass X-Ray
  Binaries}.
\newblock {\em \pasj}, 57:L17--L20, June 2005.

\bibitem{Kato2005a}
S.~{Kato}.
\newblock {Quasi-Periodic Oscillations Resonantly Induced on Spin-Induced
  Deformed-Disks of Neutron Stars}.
\newblock {\em \pasj}, 57:679--690, August 2005.

\bibitem{Kluzniak2001}
W.~{Kluzniak} and M.~A. {Abramowicz}.
\newblock {The physics of kHz QPOs---strong gravity's coupled anharmonic
  oscillators}.
\newblock {\em ArXiv Astrophysics e-prints}, May 2001.

\bibitem{Koerding2008}
E.~{K{\"o}rding}, M.~{Rupen}, C.~{Knigge}, R.~{Fender}, V.~{Dhawan},
  M.~{Templeton}, and T.~{Muxlow}.
\newblock {A Transient Radio Jet in an Erupting Dwarf Nova}.
\newblock {\em Science}, 320:1318--, June 2008.

\bibitem{Koerding2007}
E.~G. {K{\"o}rding}.
\newblock {Common disc-jet coupling in accreting objects}.
\newblock {\em \apss}, 311:143--147, October 2007.

\bibitem{Koerding2006}
E.~G. {K{\"o}rding}, R.~P. {Fender}, and S.~{Migliari}.
\newblock {Jet-dominated advective systems: radio and X-ray luminosity
  dependence on the accretion rate}.
\newblock {\em \mnras}, 369:1451--1458, July 2006.

\bibitem{Kotani2000a}
T.~{Kotani}, K.~{Ebisawa}, T.~{Dotani}, H.~{Inoue}, F.~{Nagase}, Y.~{Tanaka},
  and Y.~{Ueda}.
\newblock {ASCA Observations of the Absorption Line Features from the
  Superluminal Jet Source GRS 1915+105}.
\newblock {\em \apj}, 539:413--423, August 2000.

\bibitem{Kuulkers2006}
E.~{Kuulkers}, A.~{Norton}, A.~{Schwope}, and B.~{Warner}.
\newblock {\em {X-rays from cataclysmic variables}}, pages 421--460.
\newblock April 2006.

\bibitem{Kylafis2015}
N.~D. {Kylafis} and T.~M. {Belloni}.
\newblock {Accretion and ejection in black-hole X-ray transients}.
\newblock {\em \aap}, 574:A133, February 2015.

\bibitem{Kylafis2012}
N.~D. {Kylafis}, I.~{Contopoulos}, D.~{Kazanas}, and D.~M. {Christodoulou}.
\newblock {Formation and destruction of jets in X-ray binaries}.
\newblock {\em \aap}, 538:A5, February 2012.

\bibitem{Lee2002}
J.~C. {Lee}, C.~S. {Reynolds}, R.~{Remillard}, N.~S. {Schulz}, E.~G.
  {Blackman}, and A.~C. {Fabian}.
\newblock {High-Resolution Chandra HETGS and Rossi X-Ray Timing Explorer
  Observations of GRS 1915+105: A Hot Disk Atmosphere and Cold Gas Enriched in
  Iron and Silicon}.
\newblock {\em \apj}, 567:1102--1111, March 2002.

\bibitem{Maccarone2003}
T.~J. {Maccarone} and P.~S. {Coppi}.
\newblock {Hysteresis in the light curves of soft X-ray transients}.
\newblock {\em \mnras}, 338:189--196, January 2003.

\bibitem{Malzac2005}
J.~{Malzac}, A.~M. {Dumont}, and M.~{Mouchet}.
\newblock {Full radiative coupling in two-phase models for accreting black
  holes}.
\newblock {\em \aap}, 430:761--769, February 2005.

\bibitem{Markoff2010}
S.~{Markoff}.
\newblock {From Multiwavelength to Mass Scaling: Accretion and Ejection in
  Microquasars and AGN}.
\newblock In {T.~Belloni}, editor, {\em Lecture Notes in Physics, Berlin
  Springer Verlag}, volume 794 of {\em Lecture Notes in Physics, Berlin
  Springer Verlag}, pages 143--+, March 2010.

\bibitem{Markoff2001}
S.~{Markoff}, H.~{Falcke}, and R.~{Fender}.
\newblock {A jet model for the broadband spectrum of XTE J1118+480. Synchrotron
  emission from radio to X-rays in the Low/Hard spectral state}.
\newblock {\em \aap}, 372:L25--L28, June 2001.

\bibitem{Markwardt2005}
C.~B. {Markwardt} and J.~H. {Swank}.
\newblock {New Outburst of GRO J1655-40?}
\newblock {\em The Astronomer's Telegram}, 414:1, February 2005.

\bibitem{McClintock1986a}
J.~E. {McClintock} and R.~A. {Remillard}.
\newblock {The black hole binary A0620-00}.
\newblock {\em \apj}, 308:110--122, September 1986.

\bibitem{McHardy2010}
I.~{McHardy}.
\newblock {X-Ray Variability of AGN and Relationship to Galactic Black Hole
  Binary Systems}.
\newblock In T.~{Belloni}, editor, {\em Lecture Notes in Physics, Berlin
  Springer Verlag}, volume 794 of {\em Lecture Notes in Physics, Berlin
  Springer Verlag}, page 203, March 2010.

\bibitem{McHardy2007}
I.~M. {McHardy}, P.~{Ar{\'e}valo}, P.~{Uttley}, I.~E. {Papadakis}, D.~P.
  {Summons}, W.~{Brinkmann}, and M.~J. {Page}.
\newblock {Discovery of multiple Lorentzian components in the X-ray timing
  properties of the Narrow Line Seyfert 1 Ark 564}.
\newblock {\em \mnras}, 382:985--994, December 2007.

\bibitem{McHardy2005}
I.~M. {McHardy}, K.~F. {Gunn}, P.~{Uttley}, and M.~R. {Goad}.
\newblock {MCG-6-30-15: long time-scale X-ray variability, black hole mass and
  active galactic nuclei high states}.
\newblock {\em \mnras}, 359:1469--1480, June 2005.

\bibitem{Mendez2013}
M.~{M{\'e}ndez}, D.~{Altamirano}, T.~{Belloni}, and A.~{Sanna}.
\newblock {The phase lags of high-frequency quasi-periodic oscillations in four
  black hole candidates}.
\newblock {\em \mnras}, August 2013.

\bibitem{Merloni2003}
A.~{Merloni}, S.~{Heinz}, and T.~{di Matteo}.
\newblock {A Fundamental Plane of black hole activity}.
\newblock {\em \mnras}, 345:1057--1076, November 2003.

\bibitem{Meyer1994}
F.~{Meyer} and E.~{Meyer-Hofmeister}.
\newblock {Accretion disk evaporation by a coronal siphon flow}.
\newblock {\em \aap}, 288:175--182, August 1994.

\bibitem{Meyer-Hofmeister2014}
E.~{Meyer-Hofmeister} and F.~{Meyer}.
\newblock {The relation between radio and X-ray luminosity of black hole
  binaries: affected by inner cool disks?}
\newblock {\em \aap}, 562:A142, February 2014.

\bibitem{Middleton2010}
M.~{Middleton} and C.~{Done}.
\newblock {The X-ray binary analogy to the first AGN quasi-periodic
  oscillation}.
\newblock {\em \mnras}, 403:9--16, March 2010.

\bibitem{Migliari2006}
S.~{Migliari} and R.~P. {Fender}.
\newblock {Jets in neutron star X-ray binaries: a comparison with black holes}.
\newblock {\em \mnras}, 366:79--91, February 2006.

\bibitem{Miller2001}
J.~M. {Miller}, R.~{Wijnands}, J.~{Homan}, T.~{Belloni}, D.~{Pooley},
  S.~{Corbel}, C.~{Kouveliotou}, M.~{van der Klis}, and W.~H.~G. {Lewin}.
\newblock {High-Frequency Quasi-Periodic Oscillations in the 2000 Outburst of
  the Galactic Microquasar XTE J1550-564}.
\newblock {\em \apj}, 563:928--933, December 2001.

\bibitem{Miller-Jones2006}
J.~C.~A. {Miller-Jones}, R.~P. {Fender}, and E.~{Nakar}.
\newblock {Opening angles, Lorentz factors and confinement of X-ray binary
  jets}.
\newblock {\em \mnras}, 367:1432--1440, April 2006.

\bibitem{Mirabel1994}
I.~F. {Mirabel} and L.~F. {Rodr{\'{\i}}guez}.
\newblock {A superluminal source in the Galaxy}.
\newblock {\em \nat}, 371:46--48, September 1994.

\bibitem{Mirabel1992}
I.~F. {Mirabel}, L.~F. {Rodriguez}, B.~{Cordier}, J.~{Paul}, and F.~{Lebrun}.
\newblock {A double-sided radio jet from the compact Galactic Centre
  annihilator 1E1740.7-2942}.
\newblock {\em \nat}, 358:215--217, July 1992.

\bibitem{Miyamoto1993}
S.~{Miyamoto}, S.~{Iga}, S.~{Kitamoto}, and Y.~{Kamado}.
\newblock {Another canonical time variation of X-rays from black hole
  candidates in the very high flare state?}
\newblock {\em \apjl}, 403:L39--L42, January 1993.

\bibitem{Miyamoto1991}
S.~{Miyamoto} and S.~{Kitamoto}.
\newblock {A jet model for a very high state of GX 339 - 4}.
\newblock {\em \apj}, 374:741--743, June 1991.

\bibitem{Miyamoto1995}
S.~{Miyamoto}, S.~{Kitamoto}, K.~{Hayashida}, and W.~{Egoshi}.
\newblock {Large hysteretic behavior of stellar black hole candidate X-ray
  binaries}.
\newblock {\em \apjl}, 442:L13--L16, March 1995.

\bibitem{Miyamoto1992}
S.~{Miyamoto}, S.~{Kitamoto}, S.~{Iga}, H.~{Negoro}, and K.~{Terada}.
\newblock {Canonical time variations of X-rays from black hole candidates in
  the low-intensity state}.
\newblock {\em \apjl}, 391:L21--L24, May 1992.

\bibitem{Morgan1997}
E.~H. {Morgan}, R.~A. {Remillard}, and J.~{Greiner}.
\newblock {RXTE Observations of QPOs in the Black Hole Candidate GRS 1915+105}.
\newblock {\em \apj}, 482:993--+, June 1997.

\bibitem{Motta2009a}
S.~{Motta}, T.~{Belloni}, and J.~{Homan}.
\newblock {The evolution of the high-energy cut-off in the X-ray spectrum of GX
  339-4 across a hard-to-soft transition}.
\newblock {\em \mnras}, 400:1603--1612, December 2009.

\bibitem{Motta2012}
S.~{Motta}, J.~{Homan}, T.~{Mu{\~n}oz Darias}, P.~{Casella}, T.~M. {Belloni},
  B.~{Hiemstra}, and M.~{M{\'e}ndez}.
\newblock {Discovery of two simultaneous non-harmonically related
  quasi-periodic oscillations in the 2005 outburst of the black hole binary GRO
  J1655-40}.
\newblock {\em \mnras}, 427:595--606, November 2012.

\bibitem{Motta2011}
S.~{Motta}, T.~{Mu{\~n}oz-Darias}, P.~{Casella}, T.~{Belloni}, and J.~{Homan}.
\newblock {Low-frequency oscillations in black holes: a spectral-timing
  approach to the case of GX 339-4}.
\newblock {\em \mnras}, 418:2292--2307, December 2011.

\bibitem{Motta2011a}
S.~{Motta}, T.~{Mu{\~n}oz-Darias}, P.~{Casella}, T.~{Belloni}, and J.~{Homan}.
\newblock {Low-frequency oscillations in black holes: a spectral-timing
  approach to the case of GX 339-4}.
\newblock {\em \mnras}, 418:2292--2307, December 2011.

\bibitem{Motta2014}
S.~E. {Motta}, T.~M. {Belloni}, L.~{Stella}, T.~{Mu{\~n}oz-Darias}, and
  R.~{Fender}.
\newblock {Precise mass and spin measurements for a stellar-mass black hole
  through X-ray timing: the case of GRO J1655-40}.
\newblock {\em \mnras}, 437:2554--2565, January 2014.

\bibitem{Motta2015}
S.~E. {Motta}, P.~{Casella}, M.~{Henze}, T.~{Mu{\~n}oz-Darias}, A.~{Sanna},
  R.~{Fender}, and T.~{Belloni}.
\newblock {Geometrical constraints on the origin of timing signals from black
  holes}.
\newblock {\em \mnras}, 447:2059--2072, February 2015.

\bibitem{Motta2014a}
S.~E. {Motta}, T.~{Mu{\~n}oz-Darias}, A.~{Sanna}, R.~{Fender}, T.~{Belloni},
  and L.~{Stella}.
\newblock {Black hole spin measurements through the relativistic precession
  model: XTE J1550-564}.
\newblock {\em \mnras}, January 2014.

\bibitem{Munoz-Darias2013}
T.~{Mu{\~n}oz-Darias}, M.~{Coriat}, D.~S. {Plant}, G.~{Ponti}, R.~P. {Fender},
  and R.~J.~H. {Dunn}.
\newblock {Inclination and relativistic effects in the outburst evolution of
  black hole transients}.
\newblock {\em \mnras}, 432:1330--1337, June 2013.

\bibitem{Munoz-Darias2014}
T.~{Mu{\~n}oz-Darias}, R.~P. {Fender}, S.~E. {Motta}, and T.~M. {Belloni}.
\newblock {Black hole-like hysteresis and accretion states in neutron star
  low-mass X-ray binaries}.
\newblock {\em \mnras}, 443:3270--3283, October 2014.

\bibitem{Munoz-Darias2011}
T.~{Mu{\~n}oz-Darias}, S.~{Motta}, and T.~M. {Belloni}.
\newblock {Fast variability as a tracer of accretion regimes in black hole
  transients}.
\newblock {\em \mnras}, 410:679--684, January 2011.

\bibitem{Mucciarelli2006}
P.~{Mucciarelli}, P.~{Casella}, T.~{Belloni}, L.~{Zampieri}, and P.~{Ranalli}.
\newblock {A variable Quasi-Periodic Oscillation in M82 X-1. Timing and
  spectral analysis of XMM-Newton and RossiXTE observations}.
\newblock {\em \mnras}, 365:1123--1130, February 2006.

\bibitem{Neilsen2012}
J.~{Neilsen} and J.~{Homan}.
\newblock {A Hybrid Magnetically/Thermally Driven Wind in the Black Hole GRO
  J1655-40?}
\newblock {\em \apj}, 750:27, May 2012.

\bibitem{Nespoli2003}
E.~{Nespoli}, T.~{Belloni}, J.~{Homan}, J.~M. {Miller}, W.~H.~G. {Lewin},
  M.~{M{\'e}ndez}, and M.~{van der Klis}.
\newblock {A transient variable 6 Hz QPO from GX 339-4}.
\newblock {\em \aap}, 412:235--240, December 2003.

\bibitem{Nowak2008}
M.~A. {Nowak}, A.~{Juett}, J.~{Homan}, Y.~{Yao}, J.~{Wilms}, N.~S. {Schulz},
  and C.~R. {Canizares}.
\newblock {Disk-dominated States of 4U 1957+11: Chandra, XMM-Newton, and RXTE
  Observations of Ostensibly the Most Rapidly Spinning Galactic Black Hole}.
\newblock {\em \apj}, 689:1199--1214, December 2008.

\bibitem{Olive1998}
J.~F. {Olive}, D.~{Barret}, L.~{Boirin}, J.~E. {Grindlay}, J.~H. {Swank}, and
  A.~P. {Smale}.
\newblock {RXTE observation of the X-ray burster 1E 1724-3045. I. Timing study
  of the persistent X-ray emission with the PCA}.
\newblock {\em \aap}, 333:942--951, May 1998.

\bibitem{Oosterbroek1997}
T.~{Oosterbroek}, M.~{van der Klis}, J.~{van Paradijs}, B.~{Vaughan},
  R.~{Rutledge}, W.~H.~G. {Lewin}, Y.~{Tanaka}, F.~{Nagase}, T.~{Dotani},
  K.~{Mitsuda}, and S.~{Miyamoto}.
\newblock {Spectral and timing behaviour of GS 2023+338.}
\newblock {\em \aap}, 321:776--790, May 1997.

\bibitem{Parmar2002}
A.~N. {Parmar}, T.~{Oosterbroek}, L.~{Boirin}, and D.~{Lumb}.
\newblock {Discovery of narrow X-ray absorption features from the dipping
  low-mass X-ray binary X 1624-490 with XMM-Newton}.
\newblock {\em \aap}, 386:910--915, May 2002.

\bibitem{Pasham2013}
D.~R. {Pasham} and T.~E. {Strohmayer}.
\newblock {On the Nature of the mHz X-Ray Quasi-periodic Oscillations from
  Ultraluminous X-Ray Source M82 X-1: Search for Timing-Spectral Correlations}.
\newblock {\em \apj}, 771:101, July 2013.

\bibitem{Patterson1977}
J.~{Patterson}, E.~L. {Robinson}, and R.~E. {Nather}.
\newblock {Rapid and ultrarapid oscillations in RU Pegasi.}
\newblock {\em \apj}, 214:144--151, May 1977.

\bibitem{Pintore2014}
F.~{Pintore}, L.~{Zampieri}, A.~{Wolter}, and T.~{Belloni}.
\newblock {Ultraluminous X-ray sources: a deeper insight into their spectral
  evolution}.
\newblock {\em \mnras}, 439:3461--3475, April 2014.

\bibitem{Plotkin2013}
R.~M. {Plotkin}, E.~{Gallo}, and P.~G. {Jonker}.
\newblock {The X-Ray Spectral Evolution of Galactic Black Hole X-Ray Binaries
  toward Quiescence}.
\newblock {\em \apj}, 773:59, August 2013.

\bibitem{Plotkin2015}
R.~M. {Plotkin}, E.~{Gallo}, S.~{Markoff}, J.~{Homan}, P.~G. {Jonker}, J.~C.~A.
  {Miller-Jones}, D.~M. {Russell}, and S.~{Drappeau}.
\newblock {Constraints on relativistic jets in quiescent black hole X-ray
  binaries from broad-band spectral modelling}.
\newblock {\em \mnras}, 446:4098--4111, February 2015.

\bibitem{Ponti2012}
G.~{Ponti}, R.~P. {Fender}, M.~C. {Begelman}, R.~J.~H. {Dunn}, J.~{Neilsen},
  and M.~{Coriat}.
\newblock {Ubiquitous equatorial accretion disc winds in black hole soft
  states}.
\newblock {\em \mnras}, 422:L11, May 2012.

\bibitem{Ponti2014}
G.~{Ponti}, T.~{Mu{\~n}oz-Darias}, and R.~P. {Fender}.
\newblock {A connection between accretion state and Fe K absorption in an
  accreting neutron star: black hole-like soft-state winds?}
\newblock {\em \mnras}, 444:1829--1834, October 2014.

\bibitem{Poutanen1997}
J.~{Poutanen}, J.~H. {Krolik}, and F.~{Ryde}.
\newblock {The nature of spectral transitions in accreting black holes - The
  case of CYG X-1}.
\newblock {\em \mnras}, 292:L21--L25, November 1997.

\bibitem{Proga2002}
D.~{Proga} and T.~R. {Kallman}.
\newblock {On the Role of the Ultraviolet and X-Ray Radiation in Driving a Disk
  Wind in X-Ray Binaries}.
\newblock {\em \apj}, 565:455--470, January 2002.

\bibitem{Psaltis2008}
D.~{Psaltis}.
\newblock {Probes and Tests of Strong-Field Gravity with Observations in the
  Electromagnetic Spectrum}.
\newblock {\em Living Reviews in Relativity}, 11:9, November 2008.

\bibitem{Psaltis1999}
D.~{Psaltis}, T.~{Belloni}, and M.~{van der Klis}.
\newblock {Correlations in Quasi-periodic Oscillation and Noise Frequencies
  among Neutron Star and Black Hole X-Ray Binaries}.
\newblock {\em \apj}, 520:262--270, July 1999.

\bibitem{Pudritz2007}
R.~E. {Pudritz}, R.~{Ouyed}, C.~{Fendt}, and A.~{Brandenburg}.
\newblock {Disk Winds, Jets, and Outflows: Theoretical and Computational
  Foundations}.
\newblock {\em Protostars and Planets V}, pages 277--294, 2007.

\bibitem{Reig2003}
P.~{Reig}, T.~{Belloni}, and M.~{van der Klis}.
\newblock {Does GRS 1915+105 exhibit ``canonical'' black-hole states?}
\newblock {\em \aap}, 412:229--233, December 2003.

\bibitem{Reig2000}
P.~{Reig}, T.~{Belloni}, M.~{van der Klis}, M.~{M{\'e}ndez}, N.~D. {Kylafis},
  and E.~C. {Ford}.
\newblock {Phase Lag Variability Associated with the 0.5-10 HZ Quasi-Periodic
  Oscillations in GRS 1915+105}.
\newblock {\em \apj}, 541:883--888, October 2000.

\bibitem{Remillard2002}
R.~A. {Remillard}, M.~P. {Muno}, J.~E. {McClintock}, and J.~A. {Orosz}.
\newblock {Evidence for Harmonic Relationships in the High-Frequency
  Quasi-periodic Oscillations of XTE J1550-564 and GRO J1655-40}.
\newblock {\em \apj}, 580:1030--1042, December 2002.

\bibitem{Remillard2006}
R.A. {Remillard} and J.E. {McClintock}.
\newblock {X-Ray Properties of Black-Hole Binaries}.
\newblock {\em \araa}, 44:49--92, September 2006.

\bibitem{Reynolds2012}
C.~S. {Reynolds}.
\newblock {Constraints on Compton-thick Winds from Black Hole Accretion Disks:
  Can We See the Inner Disk?}
\newblock {\em \apjl}, 759:L15, November 2012.

\bibitem{Ruhlen2011}
L.~{Ruhlen}, D.~M. {Smith}, and J.~H. {Swank}.
\newblock {The Nature and Cause of Spectral Variability in LMC X-1}.
\newblock {\em \apj}, 742:75, December 2011.

\bibitem{Russell2006}
D.~M. {Russell}, R.~P. {Fender}, R.~I. {Hynes}, C.~{Brocksopp}, J.~{Homan},
  P.~G. {Jonker}, and M.~M. {Buxton}.
\newblock {Global optical/infrared-X-ray correlations in X-ray binaries:
  quantifying disc and jet contributions}.
\newblock {\em \mnras}, 371:1334--1350, September 2006.

\bibitem{Russell2011a}
D.~M. {Russell}, J.~C.~A. {Miller-Jones}, T.~J. {Maccarone}, Y.~J. {Yang},
  R.~P. {Fender}, and F.~{Lewis}.
\newblock {Testing the Jet Quenching Paradigm with an Ultradeep Observation of
  a Steadily Soft State Black Hole}.
\newblock {\em \apjl}, 739:L19, September 2011.

\bibitem{Russell2015}
T.~D. {Russell}, J.~C.~A. {Miller-Jones}, P.~A. {Curran}, R.~{Soria},
  D.~{Altamirano}, S.~{Corbel}, M.~{Coriat}, A.~{Moin}, D.~M. {Russell}, G.~R.
  {Sivakoff}, T.~J. {Slaven-Blair}, T.~M. {Belloni}, R.~P. {Fender},
  S.~{Heinz}, P.~G. {Jonker}, H.~A. {Krimm}, E.~G. {K{\"o}rding}, D.~{Maitra},
  S.~{Markoff}, M.~{Middleton}, S.~{Migliari}, R.~A. {Remillard}, M.~P.
  {Rupen}, C.~L. {Sarazin}, A.~J. {Tetarenko}, M.~A.~P. {Torres}, V.~{Tudose},
  and A.~K. {Tzioumis}.
\newblock {Radio monitoring of the hard state jets in the 2011 outburst of MAXI
  J1836-194}.
\newblock {\em \mnras}, 450:1745--1759, June 2015.

\bibitem{Servillat2011}
M.~{Servillat}, S.~A. {Farrell}, D.~{Lin}, O.~{Godet}, D.~{Barret}, and N.~A.
  {Webb}.
\newblock {X-Ray Variability and Hardness of ESO 243-49 HLX-1: Clear Evidence
  for Spectral State Transitions}.
\newblock {\em \apj}, 743:6, December 2011.

\bibitem{Shakura1973}
N.~I. {Shakura} and R.~A. {Sunyaev}.
\newblock {Black holes in binary systems. Observational appearance.}
\newblock {\em \aap}, 24:337--355, 1973.

\bibitem{Shidatsu2013}
M.~{Shidatsu}, Y.~{Ueda}, S.~{Nakahira}, C.~{Done}, K.~{Morihana},
  M.~{Sugizaki}, T.~{Mihara}, T.~{Hori}, H.~{Negoro}, N.~{Kawai}, K.~{Yamaoka},
  K.~{Ebisawa}, M.~{Matsuoka}, M.~{Serino}, T.~{Yoshikawa}, T.~{Nagayama}, and
  N.~{Matsunaga}.
\newblock {The Accretion Disk and Ionized Absorber of the 9.7 hr Dipping Black
  Hole Binary MAXI J1305-704}.
\newblock {\em \apj}, 779:26, December 2013.

\bibitem{Smale2012}
A.~P. {Smale} and P.~T. {Boyd}.
\newblock {Anomalous Low States and Long-term Variability in the Black Hole
  Binary LMC X-3}.
\newblock {\em \apj}, 756:146, September 2012.

\bibitem{soleri2008}
P.~{Soleri}, T.~{Belloni}, and P.~{Casella}.
\newblock {A transient low-frequency quasi-periodic oscillation from the black
  hole binary GRS 1915+105}.
\newblock {\em \mnras}, 383:1089--1102, January 2008.

\bibitem{Soleri2011}
P.~{Soleri} and R.~{Fender}.
\newblock {On the nature of the 'radio-quiet' black hole binaries}.
\newblock {\em \mnras}, 413:2269--2280, May 2011.

\bibitem{Soleri2013}
P.~{Soleri}, T.~{Mu{\~n}oz-Darias}, S.~{Motta}, T.~{Belloni}, P.~{Casella},
  M.~{M{\'e}ndez}, D.~{Altamirano}, M.~{Linares}, R.~{Wijnands}, R.~{Fender},
  and M.~{van der Klis}.
\newblock {A complex state transition from the black hole candidate Swift
  J1753.5-0127}.
\newblock {\em \mnras}, 429:1244--1257, February 2013.

\bibitem{Soria2011a}
R.~{Soria}.
\newblock {Hard and soft spectral states of ULXs}.
\newblock {\em Astronomische Nachrichten}, 332:330, May 2011.

\bibitem{Soria2011}
R.~{Soria}, J.~W. {Broderick}, J.~{Hao}, D.~C. {Hannikainen}, M.~{Mehdipour},
  K.~{Pottschmidt}, and S.-N. {Zhang}.
\newblock {Accretion states of the Galactic microquasar GRS 1758-258}.
\newblock {\em \mnras}, 415:410--424, July 2011.

\bibitem{Stella1998}
L.~{Stella} and M.~{Vietri}.
\newblock {Lense-Thirring Precession and Quasi-periodic Oscillations in
  Low-Mass X-Ray Binaries}.
\newblock {\em \apjl}, 492:L59+, January 1998.

\bibitem{Stella1999a}
L.~{Stella} and M.~{Vietri}.
\newblock {kHz Quasiperiodic Oscillations in Low-Mass X-Ray Binaries as Probes
  of General Relativity in the Strong-Field Regime}.
\newblock {\em Physical Review Letters}, 82:17--20, January 1999.

\bibitem{Stella1999}
L.~{Stella}, M.~{Vietri}, and S.~M. {Morsink}.
\newblock {Correlations in the Quasi-periodic Oscillation Frequencies of
  Low-Mass X-Ray Binaries and the Relativistic Precession Model}.
\newblock {\em \apjl}, 524:L63--L66, October 1999.

\bibitem{Stiele2011}
H.~{Stiele}, S.~{Motta}, T.~{Mu{\~n}oz-Darias}, and T.~M. {Belloni}.
\newblock {Spectral properties of transitions between soft and hard state in GX
  339-4}.
\newblock {\em ArXiv e-prints}, August 2011.

\bibitem{Stirling2001}
A.~M. {Stirling}, R.~E. {Spencer}, C.~J. {de la Force}, M.~A. {Garrett}, R.~P.
  {Fender}, and R.~N. {Ogley}.
\newblock {A relativistic jet from Cygnus X-1 in the low/hard X-ray state}.
\newblock {\em \mnras}, 327:1273--1278, November 2001.

\bibitem{Strohmayer2001}
T.~E. {Strohmayer}.
\newblock {Discovery of a 450 HZ Quasi-periodic Oscillation from the
  Microquasar GRO J1655-40 with the Rossi X-Ray Timing Explorer}.
\newblock {\em \apjl}, 552:L49--L53, May 2001.

\bibitem{Strohmayer2001a}
T.~E. {Strohmayer}.
\newblock {Discovery of a Second High-Frequency Quasi-periodic Oscillation from
  the Microquasar GRS 1915+105}.
\newblock {\em \apjl}, 554:L169--L172, June 2001.

\bibitem{Strohmayer2003a}
T.~E. {Strohmayer} and R.~F. {Mushotzky}.
\newblock {Discovery of X-Ray Quasi-periodic Oscillations from an Ultraluminous
  X-Ray Source in M82: Evidence against Beaming}.
\newblock {\em \apjl}, 586:L61--L64, March 2003.

\bibitem{Tagger1999}
M.~{Tagger} and R.~{Pellat}.
\newblock {An accretion-ejection instability in magnetized disks}.
\newblock {\em \aap}, 349:1003--1016, September 1999.

\bibitem{Takizawa1997}
M.~{Takizawa}, T.~{Dotani}, K.~{Mitsuda}, E.~{Matsuba}, M.~{Ogawa}, T.~{Aoki},
  K.~{Asai}, K.~{Ebisawa}, K.~{Makishima}, S.~{Miyamoto}, S.~{Iga},
  B.~{Vaughan}, R.~E. {Rutledge}, and W.~H.~G. {Lewin}.
\newblock {Spectral and Temporal Variability in the X-Ray Flux of GS 1124-683,
  Nova MUSCAE 1991}.
\newblock {\em \apj}, 489:272--+, November 1997.

\bibitem{Tanaka1995}
Y.~{Tanaka} and W.~H.~G. {Lewin}.
\newblock {Black hole binaries.}
\newblock pages 126--174, 1995.

\bibitem{Tananbaum1972}
H.~{Tananbaum}, H.~{Gursky}, E.~{Kellogg}, R.~{Giacconi}, and C.~{Jones}.
\newblock {Observation of a Correlated X-Ray Transition in Cygnus X-1}.
\newblock {\em \apjl}, 177:L5+, October 1972.

\bibitem{Titarchuk2004}
L.~{Titarchuk} and R.~{Fiorito}.
\newblock {Spectral Index and Quasi-Periodic Oscillation Frequency Correlation
  in Black Hole Sources: Observational Evidence of Two Phases and Phase
  Transition in Black Holes}.
\newblock {\em \apj}, 612:988--999, September 2004.

\bibitem{Ueda1998}
Y.~{Ueda}, H.~{Inoue}, Y.~{Tanaka}, K.~{Ebisawa}, F.~{Nagase}, T.~{Kotani}, and
  N.~{Gehrels}.
\newblock {Detection of Absorption-Line Features in the X-Ray Spectra of the
  Galactic Superluminal Source GRO J1655-40}.
\newblock {\em \apj}, 492:782--787, January 1998.

\bibitem{VdK1989}
M.~{van der Klis}.
\newblock {Quasi-periodic oscillations and noise in low-mass X-ray binaries}.
\newblock {\em \araa}, 27:517--553, 1989.

\bibitem{VDK2004}
M.~{van der Klis}.
\newblock {Challenges in X-ray binary timing: current and future}.
\newblock {\em Advances in Space Research}, 34:2646--2656, 2004.

\bibitem{VDK2006}
M.~{van der Klis}.
\newblock {Overview of QPOs in neutron-star low-mass X-ray binaries}.
\newblock {\em Advances in Space Research}, 38:2675--2679, 2006.

\bibitem{Varni`ere2002}
P.~{Varni{\`e}re} and M.~{Tagger}.
\newblock {Accretion-Ejection Instability in magnetized disks: Feeding the
  corona with Alfv{\'e}n waves}.
\newblock {\em \aap}, 394:329--338, October 2002.

\bibitem{Varni`ere2012}
P.~{Varni{\`e}re}, M.~{Tagger}, and J.~{Rodriguez}.
\newblock {A possible interpretation for the apparent differences in LFQPO
  types in microquasars}.
\newblock {\em \aap}, 545:A40, September 2012.

\bibitem{Webb2012}
N.~{Webb}, D.~{Cseh}, E.~{Lenc}, O.~{Godet}, D.~{Barret}, S.~{Corbel},
  S.~{Farrell}, R.~{Fender}, N.~{Gehrels}, and I.~{Heywood}.
\newblock {Radio Detections During Two State Transitions of the
  Intermediate-Mass Black Hole HLX-1}.
\newblock {\em Science}, 337:554--, August 2012.

\bibitem{Wijnands1999}
R.~{Wijnands}, J.~{Homan}, and M.~{van der Klis}.
\newblock {The Complex Phase-Lag Behavior of the 3-12 HZ Quasi-Periodic
  Oscillations during the Very High State of XTE J1550-564}.
\newblock {\em \apjl}, 526:L33--L36, November 1999.

\bibitem{Wijnands1999a}
R.~{Wijnands} and M.~{van der Klis}.
\newblock {The Broadband Power Spectra of X-Ray Binaries}.
\newblock {\em \apj}, 514:939--944, April 1999.

\bibitem{Wilms2007}
J.~{Wilms}, K.~{Pottschmidt}, G.~G. {Pooley}, S.~{Markoff}, M.~A. {Nowak},
  I.~{Kreykenbohm}, and R.~E. {Rothschild}.
\newblock {Correlated Radio-X-Ray Variability of Galactic Black Holes: A
  Radio-X-Ray Flare in Cygnus X-1}.
\newblock {\em \apjl}, 663:L97--L100, July 2007.

\bibitem{Wu2008}
Q.~{Wu} and M.~{Gu}.
\newblock {The X-Ray Spectral Evolution in X-Ray Binaries and Its Application
  to Constrain the Black Hole Mass of Ultraluminous X-Ray Sources}.
\newblock {\em \apj}, 682:212--217, July 2008.

\bibitem{Wu2010a}
Y.~X. {Wu}, W.~{Yu}, Z.~{Yan}, L.~{Sun}, and T.~P. {Li}.
\newblock {On the relation of hard X-ray peak flux and outburst waiting time in
  the black hole transient GX 339-4}.
\newblock {\em \aap}, 512:A32, March 2010.

\bibitem{Zdziarski2003}
A.~A. {Zdziarski}, P.~{Lubi{\'n}ski}, M.~{Gilfanov}, and M.~{Revnivtsev}.
\newblock {Correlations between X-ray and radio spectral properties of
  accreting black holes}.
\newblock {\em \mnras}, 342:355--372, June 2003.

\end{thebibliography}
